\documentclass[fleqn,usenatbib]{mnras}

\usepackage{newtxtext,newtxmath}

\usepackage[T1]{fontenc}

\DeclareRobustCommand{\VAN}[3]{#2}
\let\VANthebibliography\thebibliography
\def\thebibliography{\DeclareRobustCommand{\VAN}[3]{##3}\VANthebibliography}

\usepackage{graphicx}	
\usepackage{amsmath}	

\usepackage{amssymb}	

\title[Fraction of early-type galaxies at $z<2$]{Evolution of the early-type fraction in massive galaxies at $z<2$: how did early-type morphology form?}

\author[M. Kajisawa]{
Masaru Kajisawa,$^{1,2}$\thanks{E-mail: kajisawa.masaru.mk@ehime-u.ac.jp}
\\
$^{1}$Graduate School of Science and Engineering, Ehime University, Bunkyo-cho, Matsuyama 790-8577, Japan\\
$^{2}$Research Center for Space and Cosmic Evolution, Premier Institute Advanced Studies, Ehime University, Bunkyo-cho, Matsuyama 790-8577, Japan
}

\date{Accepted XXX. Received YYY; in original form ZZZ}

\pubyear{2026}

\begin{document}
\label{firstpage}
\pagerange{\pageref{firstpage}--\pageref{lastpage}}
\maketitle

\begin{abstract}
  Using $JWST$/NIRCam data over a 0.28 deg$^{2}$ area from COSMOS-Web
    survey, together with $HST$/ACS data, 
  we investigate early-type fraction of massive galaxies
  with $M_{\rm star}>10^{10.5}M_{\odot}$ at $0.2<z<2.0$, 
  and explore the formation of their early-type morphology. 
  We measure concentration index $C$ ($=R_{80}/R_{20}$)
  and asymmetry index $A$,
  and select early-type galaxies with $C>C_{n=2.5}$ and $A_{\rm cor}<0.2$.
  Here $C_{n=2.5}$ is the concentration expected for a S{\'e}rsic profile with
  $n=2.5$ under the spatial resolution and depth of the data, and $A_{\rm cor}$ is
the asymmetry corrected for resolution effects.
The fraction of early-type galaxies with 
$M_{\rm star}>10^{11}M_{\odot}$ ($=10^{10.5}$--$10^{11}M_{\odot}$) decreases with
increasing redshift from $\sim70$\% ($\sim40$--60\%) at $z\sim0.3$
to $\sim20$--25\% ($\sim15$--25\%) at $z\sim1.8$.
We also examine the evolution of their $R_{20}$ and $R_{80}$, which
enclose 20\% and 80\% of the total flux of the galaxy, respectively. 
The median $R_{80}$ shows strong mass dependence and significant redshift
evolution, whereas the median $R_{20}$ shows little dependence 
on either stellar mass or redshift.    
In contrast, morphological differences are more pronounced 
 in $R_{20}$ than in $R_{80}$:  
the median $R_{20}$ of early-type galaxies is smaller than that of 
late-type and irregular galaxies by 0.25--0.45 and 0.3--0.6 dex, respectively.
The median SSFR of sample galaxies strongly correlates with $R_{20}$, 
and early-type galaxies have lower SSFRs by $\sim1$ dex.
We further find that early-type galaxies at $z\gtrsim1.3$ have
younger mass-weighted
stellar ages of $t_{\rm mw}\lesssim2$ Gyr than late-type and irregular ones.
Their SSFRs, $t_{\rm mw}$, and morphological properties suggest that
these high-$z$ early-type galaxies experienced  
rapid formation of a dense stellar core through starburst,
followed by quenching of star formation, 
and subsequently resumed star formation $\sim1$--2 Gyr later.
\end{abstract}




\section{Introduction}

Early-type  galaxies such as elliptical and S0 galaxies
dominate the massive end of galaxy population 
and significantly contribute to the total stellar mass budget
in the present universe 
(\citealp{kau03}; \citealp{dri07}; \citealp{kel14}).
Morphology of these early-type galaxies is characterised by
featureless light distribution and 
centrally concentrated surface brightness profile such as the 
de Vaucouleurs' law \citep{dev48}.
They have little star formation, old mean stellar ages, 
and a small amount of cold gas, which are different from late-type galaxies, 
and the morphology of galaxies is correlated with their various physical
properties (e.g., \citealp{rob94}; \citealp{blu19}).
Thus how early-type morphology was formed and
how their stellar mass was assembled  
are important issues for understanding the galaxy formation and evolution.

Investigating morphological properties of galaxies at various redshifts
is essential to understand the morphological evolution of galaxies over
the cosmic time.
The high resolution imaging capability of {\it Hubble Space Telescope} ($HST$)
has enabled to statistically study morphology of galaxies
at intermediate to high redshifts, 
and various imaging surveys with $HST$ such as HDF, GOODS, HUDF, COSMOS,
and so on have been carried out to reveal the morphological evolution of
galaxies (\citealp{wil96}; \citealp{gia04}; \citealp{bec06}; \citealp{sco07a}).
In analyses with the $HST$ data, 
distant early-type galaxies have been selected by non-parametric morphological
parameters such as concentration and asymmetry, or parametric fitting of
their surface brightness profiles with S{\'e}rsic's law \citep{ser68}
as well as visual classification (\citealp{con14} for review).
Studies with optical $HST$ data revealed that fraction of early-type
galaxies evolves only mildly at $z \lesssim 1$, 
while that of relatively bright irregular/peculiar galaxies increases
with increasing redshift (e.g., \citealp{bri98}; \citealp{bri00};
\citealp{van00}).
Furthermore, near-infrared (NIR) instruments such as NICMOS and WFC3
on board $HST$ have allowed to study rest-frame optical morphologies
of galaxies up to $z\sim$ 2--3, 
and studies with the NIR imaging data found that
the fraction of early-type galaxies significantly decreases with
redshift at $z>1$
(\citealp{kaj01}; \citealp{kaj05}; \citealp{bun05}; \citealp{wuy11}; 
\citealp{bel12}; \citealp{bui13}; \citealp{tal14}; \citealp{bru14};
\citealp{dav17}).
Previous studies with the similar NIR data also investigated 
the concentration and  S{\'e}rsic indices of massive galaxies at $z>1$,
and found that mean/median values of these indices similarly decreases with
increasing redshift (\citealp{van10}; 
\citealp{pat13}; \citealp{van13};
\citealp{mor14}; \citealp{pap15}; \citealp{gu19}; 
\citealp{ren24}).
These results suggest that a significant fraction of early-type galaxies
appeared between $z \sim 3$ and $z \sim $ 0.5--1,
and the early-type morphology could be formed preferentially at these epochs.

The correlation between morphology/structure and star formation activity
of galaxies, 
namely, lower specific star formation rates in those with early-type
morphology, has been observed up to $z \sim $ 2--3  
(e.g., \citealp{wuy11}; \citealp{bel12}; \citealp{lee13}; \citealp{bru14}).
In particular, many previous studies found that the quenching of star
formation is closely related with central surface stellar mass density
within 1 kpc, $\Sigma_{\rm 1kpc}$ (e.g., \citealp{che12}; \citealp{fan13};
\citealp{bar17}; \citealp{mos17}; \citealp{whi17}; \citealp{lee18};
\citealp{xu24}), and the correlation with $\Sigma_{\rm 1kpc}$
is clearly stronger than the known ones with 
the surface stellar mass density or inferred velocity dispersion
($=M_{\rm star}/R_{\rm e}$) within half-light (or mass) radius $R_{\rm e}$ 
(e.g., \citealp{kau03}; \citealp{bri04}; \citealp{fra08}; \citealp{omo14}).
Median specific star formation rate (SSFR) of galaxies decreases
with $\Sigma_{\rm 1kpc}$ rapidly at 
$\Sigma_{\rm 1kpc} \gtrsim 10^{9.5} M_{\odot}/{\rm kpc}^{2}$.
Such relationship suggests that the dense stellar core at the centre of
galaxies is required for the quenching of star formation,
although which mechanism(s) worked dominantly in the quenching process
is still unclear (e.g., \citealp{hop08}; \citealp{mar09}; \citealp{lap23}).
The dense stellar core and the quenching of star formation 
could affect mass assembly of these galaxies and drive 
the formation of early-type morphology.

Multi-band photometric data with high spatial resolution also have enabled to
investigate spatially resolved star formation rate and stellar mass density
of massive galaxies at $z \lesssim 3$.
Evolution of the surface stellar mass density profile of massive galaxies with
$M_{\rm star} \gtrsim 10^{11} M_{\odot}$ suggests that the stellar mass
growth occurred preferentially in their outer regions at $z \lesssim 2$ 
(\citealp{van10}; \citealp{pat13}; \citealp{mos17}).
Spatially resolved SSFR maps for those massive galaxies at $z\sim$ 1--2
also indicate the similar inside-out stellar mass growth
(e.g., \citealp{tac15}; \citealp{nel16}).
On the other hand, several studies reported that Milky Way progenitors 
that are expected to have $M_{\rm star} \sim 5 \times 10^{10} M_{\odot}$ at
$z \sim 0$ grew their stellar mass with similar rates 
over the entire galaxy at $z \lesssim 2$
(e.g., \citealp{van13}; \citealp{mor15}; \citealp{mos17}; \citealp{tan24}).
Such differences in the stellar mass assembly probably lead to the stronger
size evolution for massive (quiescent) galaxies than that for less massive
star-forming galaxies (e.g., \citealp{van14}), 
and could be related with the mass dependence of the early-type fraction.

On the other hand, the cosmological surface brightness dimming and 
the PSF smoothing effect become more severe for galaxies 
at high redshifts, and 
they could bias the results in the morphological studies mentioned above.
Missing the extended low surface brightness envelope of galaxies at $z>1$ 
could lead to underestimate of the size and/or S{\'e}rsic index (e.g.,
\citealp{man10}; \citealp{sal24}; \citealp{wan24}).
The image smearing by PSF could also cause underestimate of the
concentration/S{\'e}rsic index and asymmetry index 
for compact galaxies, which are more ubiquitous
at high redshifts (e.g., \citealp{wan24}).
Deeper and sharper imaging data are essential to precisely determine 
the evolution of early-type galaxies and reveal how their morphology formed.

Recently, {\it James Webb Space Telescope} ($JWST$) provides 
deeper and sharper NIR 
imaging data, which allow us to carry out more precise and detailed
morphological studies of galaxies at high redshifts  
(e.g., \citealp{fer23}; \citealp{orm24};
\citealp{wri24}; \citealp{lee24}; \citealp{ren24}; \citealp{kuh24};
\citealp{lec24}). 
In this paper, we investigate the early-type fraction in galaxies with 
$M_{\rm star} > 10^{10.5} M_{\odot}$ at $0.2<z<2.0$  
and how the early-type morphology formed by using 
$JWST$/NIRCam imaging data from COSMOS-Web survey \citep{cas23} over 
a 0.28 deg$^{2}$ area in the COSMOS field as well as $HST$/ACS data.
The high spatial resolution allows us to examine detailed surface 
brightness distribution of galaxies even near the centre, while   
the depth and relatively wide area of the data ensure completeness for
the early-type selection and large sample size.
With these data, we select early-type galaxies paying attention to the effects 
of the surface brightness limit and PSF smoothing.
Section 2 describes the data used in this study.
In Section 3, we describe sample selection with SED fitting and methods to
measure non-parametric morphological indices for sample galaxies.
Section 4 presents the fraction of early-type galaxies and evolution of
their physical properties.
We discuss our results and their implications in Section 5, and summarise the
results of this study in Section 6.
Throughout this paper, 
 we assume a flat universe with $\Omega_{\rm m}=0.3$, $\Omega_{\Lambda}=0.7$,
and $H_{0}=70$ km s$^{-1}$ Mpc$^{-1}$, and magnitudes are given in the AB system.

\section{Data} \label{sec:data}

In this study, we use multi-band photometry from COSMOS2020 catalogue 
\citep{wea22} to construct a sample of massive galaxies with
$M_{\rm star} > 10^{10.5} M_{\odot}$ at $0.2<z<2.0$.
\citet{wea22} provided multi-band photometry from UV to MIR 
wavelengths, namely, $GALEX$ FUV and NUV \citep{zam07}, CFHT/MegaCam $u$ and $u^{*}$ \citep{saw19}, Subaru/HSC $grizy$ \citep{aih19}, Subaru/Suprime-Cam
$Bg^{'}Vr^{'}i^{'}z^{'}z^{''}$ \citep{tan07} and 12 intermediate and 2 narrow bands
\citep{tan15}, VISTA/VIRCAM $YJHK_{s}$ and $NB119$ \citep{mcc12}, and
$Spitzer$/IRAC ch1--4 (\citealp{ash13}; \citealp{ste14}; \citealp{ash15}: \citealp{ash18}), for objects in the COSMOS field \citep{sco07a}. 
Source detection was performed on an $izYJHK_{s}$-band combined image, and
aperture photometry was done on PSF-matched images with SExtractor
\citep{ber96}.
We pick up objects with $[3.6] < 26$ mag from ``CLASSIC'' catalogue of
COSMOS2020 
and use the photometry measured with 3 arcsec diameter apertures   
in the $GALEX$ NUV, MegaCam $u$
and $u^{*}$, HSC $grizy$, Suprime-Cam $BVi^{'}z^{''}$ and 12 intermediate, 
VIRCAM $YJHK_{s}$, and IRAC ch1--4 bands to construct our sample
(Sections 3.1 and 3.2). 
The photometric offsets derived from the SED fitting with LePhare 
\citep{wea22} are applied to these fluxes. 
Furthermore, we similarly estimate additional small photometric offsets
by performing 
the SED fitting described in the next section for isolated galaxies
with spectroscopic redshifts from zCOSMOS (\citealp{lil07}; \citealp{lil09})
and LEGA-C (\citealp{van16}; \citealp{van21}) surveys,
and apply them. 
We correct these fluxes for Galactic extinction by using $E(B-V)$ value
of the Milky Way at each object position from the catalogue.

In order to measure non-parametric morphological indices for sample galaxies,
we use {\it HST}/ACS F814W-band data \citep{koe07} 
and {\it JWST}/NIRCam F115W and F150W-band
imaging data produced by \citet{zhu24}.
The F814W-band data are publicly available
COSMOS {\it HST}/ACS data version 2.0, 
which cover a 1.6 deg$^2$ area in the COSMOS field and 
have a pixel scale of 0.03 arcsec/pixel and a PSF FWHM of $\sim 0.1$ arcsec.
The $5\sigma$ limiting magnitude for point sources of the F814W-band data
is $\sim 27.2$ mag \citep{sco07b}, and the pixel-to-pixel fluctuation of
the background sky corresponds to 24.5--24.8 mag/arcsec$^{2}$.
The NIRCam data were obtained in COSMOS-Web survey \citep{cas23}, 
which covers a 0.54 deg$^2$ area with NIRCam.
\citet{zhu24} reduced the NIRCam images over a 0.28 deg$^{2}$ area 
that observed by 2023 June.
The NIRCam data consist of 80 tiles (visits) for each band, 
and we align these images to the F814W-band data 
by matching positions of relatively
bright point sources in the images with IRAF/geotran task.
The aligned NIRCam images have a pixel scale of 0.03 arcsec/pixel and
PSF FWHMs of $\sim 0.065$ and 0.069 arcsec for F115W and F150W 
bands, respectively.
These PSF FWHMs are slightly larger than those of the reduced images
produced by \citet{zhu24} 
due to the pixel-to-pixel interpolation in the image alignment.
The average $5\sigma$ limiting magnitudes for 0.3 arcsec apertures 
are 27.40 and 27.66 mag in F115W and F150W bands, respectively.
The pixel-to-pixel fluctuations of the background sky in the F115W and
F150W bands correspond to 24.0--24.4 mag/arcsec$^{2}$ and
24.3--24.7 mag/arcsec$^{2}$, respectively.

\section{Analysis} \label{sec:ana}

\begin{figure*} 
  \includegraphics[width=2.0\columnwidth]{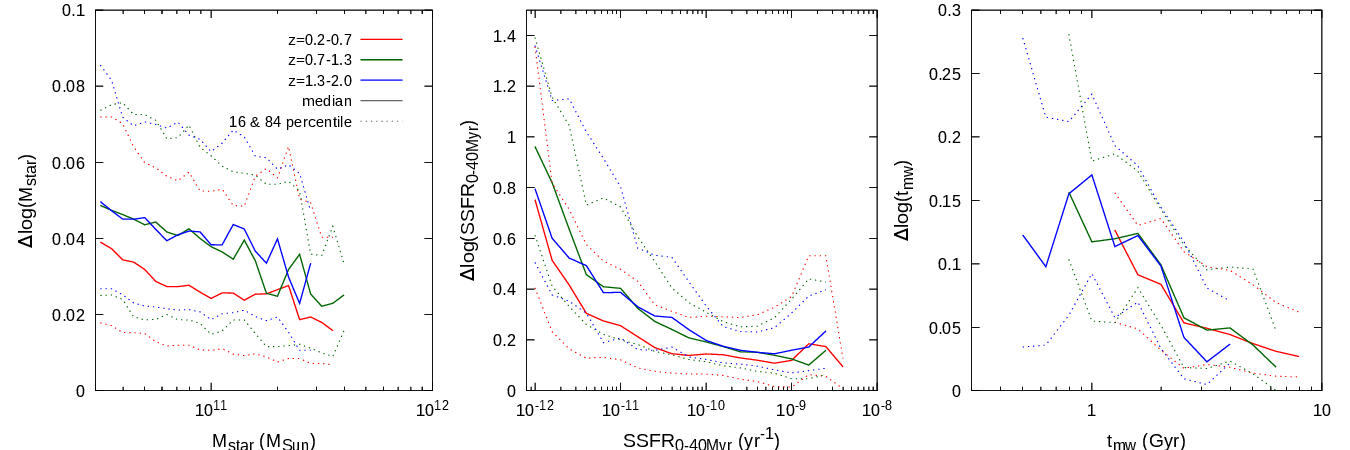}
  \caption{Uncertainty of total stellar mass, $M_{\rm star}$ (left),
    SSFR in the past 40 Myr, 
    SSFR$_{\rm 0-40Myr}$ (middle), and mass-weighted mean stellar age,
    $t_{\rm mw}$ (right) estimated in the SED fitting analysis  
    for galaxies with $M_{\rm star} > 10^{10.5} M_{\odot}$ at $0.2<z<2.0$.
    The solid lines show median errors in logarithmic scale as a function of
    the property itself, and the dotted lines represent 16 and 84 percentiles
    of the errors.
    The different colours represent the different redshift ranges. 
    \label{fig:err}}
\end{figure*}

\subsection{SED fitting} \label{sec:sed}

We fit the multi-band photometry of objects with $[3.6]<26$ mag from the
COSMOS2020 catalogue described above with population 
synthesis models of GALAXEV \citep{bru03} in order to estimate their
photometric redshift, stellar mass, SSFR, 
and mean stellar age.
We perform the SED fitting with the same methods as in \citet{him23}.
We here briefly describe the method, and refer the reader
to \citet{him23} for more details.
\citet{him23} adopt a non-parametric form of star formation history
(SFH) with constant values of SFR for several given time intervals as
free parameters, following previous studies (\citealp{toj07}; \citealp{lej17};
\citealp{cha18}).
The look-back time is divided into seven periods, namely,
0--40 Myr, 40--321 Myr, 321--1000 Myr, 1--2 Gyr, 2--4 Gyr, 4--8 Gyr, and
8--12 Gyr before observation.
We choose these look-back time periods so that model templates of the different
periods show variations in the photometric SED shape.
The model SED templates of stars formed in these periods are constructed  
under the assumptions of Chabrier IMF \citep{cha03} and
constant SFR in each period. 
We use a linear combination of the seven templates as a model SED in
the fitting. Normalisation coefficients for the seven templates, which
correspond to SFRs in these periods, are free parameters.
Those templates whose minimum ages are larger than the age of the universe
at the redshift are excluded from the fitting.
When the age of the universe enters between the minimum and maximum ages
of a template, we replace it with a new template with the maximum age
slightly younger than the age of the universe.
Non-Negative Least Squares algorithm
\citep{law74} is adopted to search the best-fit SFHs that give 
the minimum $\chi^{2}$ as in \citet{mag15},
while we use a simple full grid search for redshift, metallicity, and dust
extinction.
The templates with three stellar metallicities, namely, 0.2, 0.4, and 1.0
$Z_{\odot}$ are fitted.
If we include templates with 2.5 $Z_{\odot}$, the number of galaxies fitted
with the templates with 2.5 $Z_{\odot}$ is rather small, and results in this
study do not change.
We fix the metallicity over all the periods 
except for the youngest one, 0--40 Myr before observation.
The metallicity of the template of 0--40 Myr is independently chosen
from the same 0.2, 0.4, and 1.0 $Z_{\odot}$.
We add nebular emission only in the youngest template 
 because a contribution from 
the nebular emission is negligible in templates of the other
older periods.
We adopt the nebular continuum and hydrogen recombination lines calculated
by PANFIT (\citealp{maw16}; \citealp{maw20}), while we estimate
fluxes of other strong emission lines from empirical emission line ratios
for nearby star-forming galaxies with various gas metallicities
by assuming the same gas metallicity with the stellar one 
(\citealp{him23} for details).
For the dust extinction, we use the Calzetti law \citep{cal00} 
and attenuation curves for local star-forming galaxies with different stellar
masses, namely $10^{8.5}$--$10^{9.5} M_{\odot}$, $10^{9.5}$--$10^{10.5} M_{\odot}$,
and $10^{10.5}$--$10^{11.5} M_{\odot}$, from \citet{sal18}
to cover observed variation in 2175\AA\ bump.
We adopt different ranges of $E(B-V)$ (or $A_{V}$) for these different
attenuation curves, namely, $E(B-V) \leq 1.6$ for the Calzetti law and
$E(B-V) \leq 0.4$ for those from \citet{sal18}, 
to take account of observed correlation between 
the overall slope and $V$-band attenuation (\citealp{sal18}; \citealp{sal20}).

\begin{figure}
  \includegraphics[width=\columnwidth]{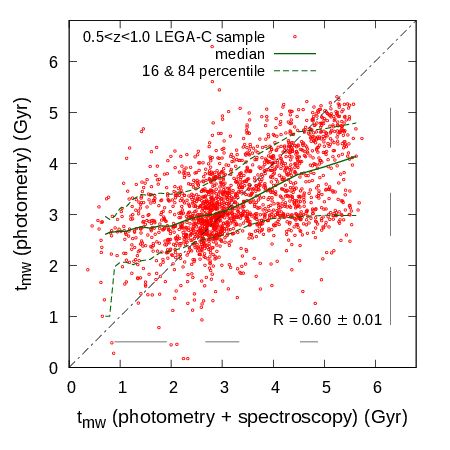}
  \caption{
    Comparison between $t_{\rm mw}$ used in this study, which is estimated
    from the photometric SED, and that estimated from  
    both the same photometric SED and optical spectrum from the
    LEGA-C survey (\citealp{van21}) for galaxies with
    $M_{\rm star} > 10^{10.5} M_{\odot}$ at $z =$ 0.5--1.0.
    The solid line shows the median values of $t_{\rm mw}$ estimated from
    only the photometric SED at a given $t_{\rm mw}$ estimated from both the
    photometric SED and LEGA-C spectrum, while the dashed lines
    represent 16 and 84 percentiles.
    The median 68\% confidence intervals of $t_{\rm mw}$ estimated from 
    the photometric SED at $t_{\rm mw} < $ 2 Gyr, 2--4 Gyr, and 
    $>$ 4 Gyr are shown on the right side of the panel, while those of
    $t_{\rm mw}$ estimated from both the photometric SED and LEGA-C spectrum are
    shown at the bottom.
    The correlation coefficient between these two mass-weighted mean ages
    is also shown.
  \label{fig:LEGAC}}
\end{figure}

In the fitting, we search the minimum $\chi^{2}$ at each redshift,
and calculate a redshift likelihood function,
$P(z) \propto \exp(-\frac{\chi^{2}(z)}{2})$,
where $\chi^{2}(z)$ is the minimum $\chi^{2}$ at each redshift.
We adopt the median of the likelihood function as a redshift of each
object (e.g., \citealp{ilb10}).
The photometric data with a central rest-frame wavelength longer than
25000 \AA\ 
are excluded from the calculation of the minimum $\chi^{2}$ at each redshift, 
because our model templates do not include the dust/PAH emission.
We check accuracy of the estimated redshifts for $\sim 4800$ galaxies
with $M_{\rm star} > 10^{10.5} M_{\odot}$ at $z<3.0$ by using spectroscopic
redshifts from various surveys (\citealp{lil07}; \citealp{lil09};
\citealp{van16}; \citealp{van21}; \citealp{sil15}; \citealp{tas17};
\citealp{has18}).
The fraction of those with $(z_{\rm phot}-z_{\rm spec})/(1+z_{\rm spec}) > 0.1$ 
is 0.2\% at $z = $0.2--1.0 and 1.4\% at $z = $1.0--2.0, and
the means and standard deviations of $(z_{\rm phot}-z_{\rm spec})/(1+z_{\rm spec})$
for galaxies with $(z_{\rm phot}-z_{\rm spec})/(1+z_{\rm spec}) < 0.1$
are $-0.007 \pm 0.013$ at $z = $0.2--1.0 and $-0.013 \pm 0.037$
at $z = $1.0--2.0.

\begin{figure*} 
  \includegraphics[width=2.0\columnwidth]{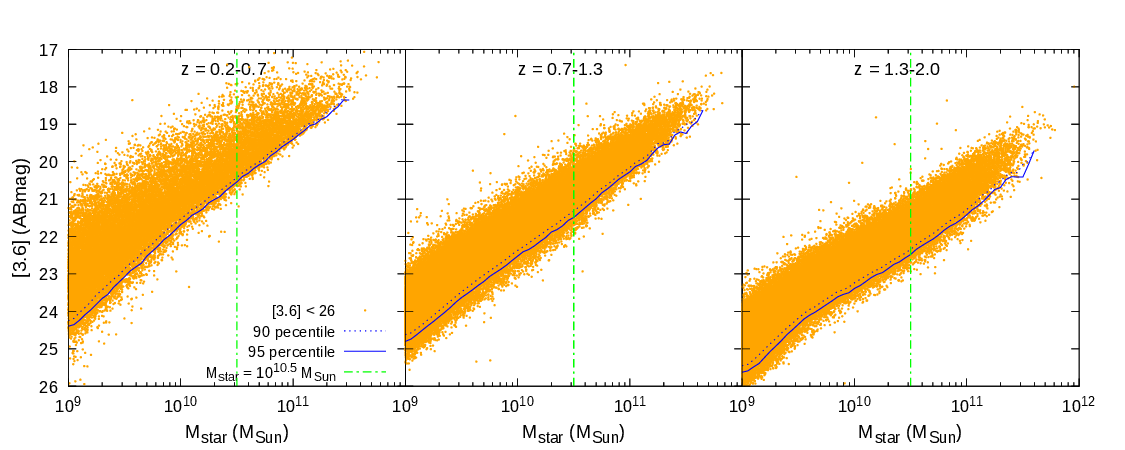}
  \caption{$[3.6]$ magnitude vs. stellar mass for galaxies with $[3.6] < 26$
    in the different redshift ranges.
    The blue dotted and solid lines represent
    90 and 95 percentiles of [3.6] magnitude in each stellar mass bin
    with a width of $\pm 0.1$ dex.
    The vertical dotted-dashed line shows the stellar mass limit of
    $M_{\rm star} = 10^{10.5} M_{\odot}$. 
    \label{fig:ch1Ms}}
\end{figure*}

In order to estimate probability distribution of derived physical
properties, namely, stellar mass, SSFR, and mean stellar age,  
we carry out Monte Carlo simulations.
We add random shifts based on photometric errors to the observed fluxes
and then perform the same SED fitting with the simulated photometries
fixing the redshift to the value described above.
In each simulation, we calculate total stellar mass, SSFR in the past
40 Myr, and mass-weighted mean stellar age $t_{\rm mw}$ of the object
from the fitted SFH.
We perform 1000 such simulations for each object and adopt the median values
and 68\% confidence intervals as the physical properties and their uncertainty,
respectively.
Figure \ref{fig:err} shows median errors of these physical properties 
in logarithmic scale as a function of the property itself for galaxies
with $M_{\rm star} > 10^{10.5} M_{\odot}$ at $0.2<z<2.0$.
One can see that stellar mass is constrained with 
$\Delta\log{M_{\rm star}} \lesssim 0.08$ dex even for galaxies 
with $M_{\rm star} \sim 10^{10.5} M_{\odot}$.
The errors in SSFR$_{\rm 0-40Myr}$ are $\lesssim 0.4$ dex 
at SSFR$_{\rm 0-40Myr} \gtrsim $10$^{-11}$--$10^{-10.5}$ yr$^{-1}$,
while the uncertainty becomes larger at lower SSFR.
Those in mass-weighted age are $\sim 0.15$ dex at $t_{\rm mw} \sim 1$ Gyr and 
decrease to $\lesssim 0.1$ dex at $t_{\rm mw} > 2$ Gyr.
The SSFR$_{\rm 0-40Myr}$ and mass-weighted mean age do not strongly
change if we adopt slightly different choices of the look-back time periods
in the SED fitting. While the mean stellar ages of very young galaxies
(e.g., $\lesssim 100$ Myr) could depend on the number and extent of the
young periods, our sample galaxies with $M_{\rm star}>10^{10.5} M_{\odot}$ at 
$0.2<z<2.0$ (see the next subsection) do not include such very young objects.

Since the physical properties estimated from photometric SEDs such as 
   mean stellar age tend to have relatively large uncertainty 
   (e.g., \citealp{con13}; \citealp{ner24}),
   we also estimate the mass-weighted mean ages of galaxies with
   $M_{\rm star} > 10^{10.5} M_{\odot}$ at $z=$ 0.5--1.0 by adding optical spectra
   from the LEGA-C survey \citep{van21} to the SED fitting analysis, and
   compare them with those estimated only from the photometric SEDs.
   The LEGA-C spectra typically have a wavelength range of
   $\sim $ 6300--8800 \AA, and we measure $i$-band fluxes on the spectra
   to match their normalisation to the photometric SEDs.
   We simultaneously fit the photometric SED and LEGA-C spectrum of these
   galaxies with the same model template set described above.
   We also carry out the similar Monte Carlo simulations to estimate
   the median values and 68\% confidence intervals of the mean stellar ages.
   In Figure \ref{fig:LEGAC}, we compare the mass-weighted mean ages estimated 
   from only the photometric SED and those
   from both the photometric and spectroscopic
   data for 1967 those galaxies.
   While the mean ages estimated from only the photometric SED are
   correlated with
   those estimated from both the photometric and spectroscopic data, there is
   a relatively large scatter at a given age estimated from the photometric
   and spectroscopic data. The correlation coefficient between them is
   $0.60 \pm 0.01$, where the uncertainty is estimated from the Monte Carlo
   simulations described above.
   In addition to the scatter, the SED fitting with only the photometric data
   tends to systematically underestimate the mean stellar ages
   at relatively old age of $t_{\rm mw} > 4$ Gyr and overestimate
   at young age of $t_{\rm mw} < 2$ Gyr. 
   These systematic effects could smear the distribution of $t_{\rm mw}$
   to some extent.
   In Section \ref{sec:ssfrage},
   we keep in mind these effects to examine the age distribution of our
   sample galaxies, and also present 
   the mean ages estimated from both the photometric SED and LEGA-C spectrum
   for these galaxies to check these effects on our results.

\begin{table*}
	\centering
	\caption{The numbers of sample galaxies with the $HST$ and $JWST$ data in the stellar mass and redshift bins. The numbers in the parentheses represent the parent sample from the COSMOS2020 catalogue over the entire COSMOS field.}
	\label{tab:sample}
	\begin{tabular}{ccccccc} 
		\hline
		redshift & 0.2--0.45 & 0.45--0.7 & 0.7--1.0 & 1.0--1.3 & 1.3--1.65 & 1.65--2.0 \\
		\hline
		$M_{\rm star} > 10^{11} M_{\odot}$ & 215 (231) & 527 (583) & 314 (1673) & 241 (1397) & 210 (1294) & 207 (1187)\\
		$M_{\rm star} = 10^{10.75}$--$10^{11} M_{\odot}$ & 411 (454) & 824 (903) & 393 (2209) & 326 (1954) & 293 (1879) & 345 (1715)\\
		$M_{\rm star} = 10^{10.5}$--$10^{10.75} M_{\odot}$ & 623 (677) & 1096 (1197) & 481 (2611) & 443 (2636) & 419 (2463) & 410 (2104)\\
		\hline
	\end{tabular}
\end{table*}

\begin{figure*} 
  \includegraphics[width=2.0\columnwidth]{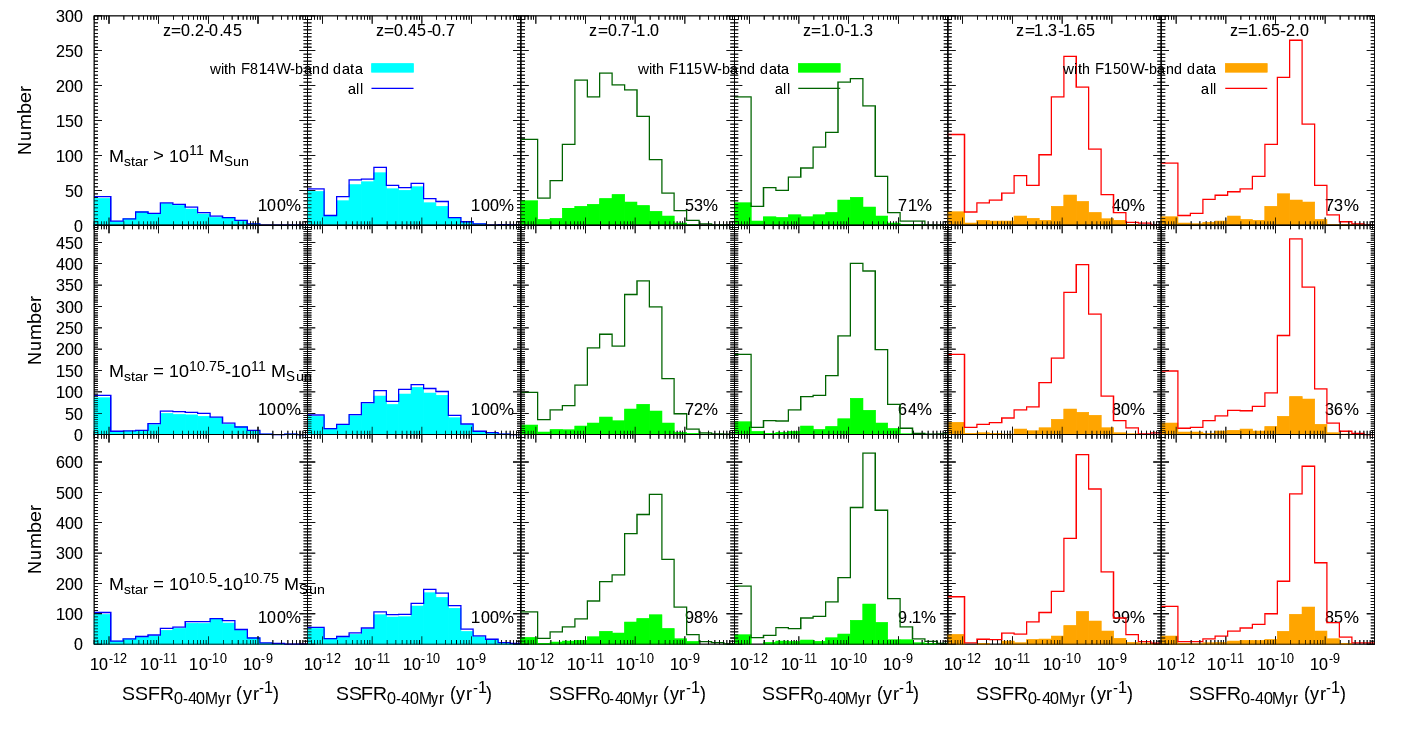}
  \caption{Distribution of SSFR$_{\rm 0-40Myr}$ for galaxies in the  
    different stellar mass and redshift bins.
    The solid histograms show those galaxies
    morphologically classified with the $HST$ and $JWST$ data, namely,  
    the F814W-band data at $0.2<z<0.7$, the F115W-band data at $0.7<z<1.3$,
    and the F150W-band data at $1.3<z<2.0$.
    The open histograms represent the parent sample from the COSMOS2020
    catalogue over the entire COSMOS field.
    Note that the leftmost bin in each panel represents all galaxies with 
    SSFR$_{\rm 0-40Myr} < 10^{-12}$ yr$^{-1}$.
    The probability that the both distributions for those galaxies
    with $HST$ and $JWST$ and the parent sample are extracted from
    the same distribution in the Kolmogorov-Smirnov test is shown at the
    bottom right of each panel.
    \label{fig:histSSFR}}
\end{figure*}

\subsection{Sample selection} \label{sec:sample}

By using photometric redshifts and stellar mass estimated with the
SED fitting mentioned above, we select 27167 galaxies with 
$M_{\rm star} > 10^{10.5} M_{\odot}$ at $0.2<z<2.0$ in the COSMOS field.
Figure \ref{fig:ch1Ms} shows distribution of [3.6] magnitude as a
function of stellar mass for galaxies 
with $[3.6] < 26$ at $0.2<z<2.0$. 
Since 90\% (95\%) of those massive galaxies
with $M_{\rm star} \sim 10^{10.5} M_{\odot}$ are $[3.6]  < 22.3$ mag ($[3.6] < 22.5$
mag) even at $z=$ 1.3--2.0, which is much brighter than the magnitude limit
of the catalogue \citep{wea22}, 
 we expect that the completeness of our sample is sufficiently high 
 although some extremely obscured galaxies with 
 $M_{\rm star} > 10^{10.5} M_{\odot}$ could be missed.

 In this study, we investigate rest-frame $V$ to $R$-band morphology
by using the F814W, F115W, and F150W-band data for those massive galaxies
at $0.2<z<0.7$, $0.7<z<1.3$, and $1.3<z<2.0$, respectively.
Out of the 27167 galaxies with $M_{\rm star} > 10^{10.5} M_{\odot}$,
the numbers of objects with the corresponding {\it HST} or {\it JWST} data
are 3696, 2198, and 1884 at $0.2<z<0.7$, $0.7<z<1.3$, and $1.3<z<2.0$,
respectively. We use total 7778 those galaxies in the following sections, 
and Table \ref{tab:sample} summarises the numbers of those galaxies
in the stellar mass and redshift bins.
Figure \ref{fig:histSSFR} shows distribution of SSFR$_{\rm 0-40Myr}$
for all massive 
galaxies and those morphologically classified with {\it HST} or {\it JWST}
data in the different stellar mass and redshift bins.
Note that the fraction of galaxies classified with the F814W-band data
at $0.2<z<0.7$ is high,
because the F814W-band data cover most of the COSMOS field. 
While the fraction is relatively small at $z>0.7$, our sample galaxies
with F115W and F150W-band data seem to represent those massive galaxies
over the entire COSMOS field in SSFR$_{\rm 0-40Myr}$. For example, 
fraction of quiescent galaxies with SSFR $< 10^{-11}$ yr$^{-1}$ in our
sample is similar with that in those massive galaxies over the entire
field in all redshift bins.
We perform the 2-sample Kolmogorov-Smirnov test for 
the SSFR distributions in the stellar mass and redshift bins,
and confirm that the two distributions are not significantly different
in all the bins.

\subsection{Morphological analysis} \label{sec:morp}

\subsubsection{Preparation}

In order to measure non-parametric morphological indices, namely,
concentration $C$, asymmetry $A$, minimum $R_{20}$,
and concentration of asymmetric features
$C_{A}$ for our sample galaxies
on the {\it HST}/ACS and {\it JWST}/NIRCam images, 
we need to identify these objects from the COSMOS2020 catalogue 
and define pixels belonging to the object on these images.
For the purpose, we run SExtractor on the ``chi-squared''
$izYJHK_{s}$-band image, on which the source detection for
the COSMOS2020 catalogue was done by \citet{wea22}, with the same parameter
set as in \citet{wea22}.
We use a segmentation image provided by the SExtractor
to define pixels belonging to each object.
We cut a 18'' $\times$ 18'' region centred on the object coordinate of 
the ACS and NIRCam data for each galaxy in order to ensure enough area
for the background sky estimate, 
and then align the segmentation image to these object images.

While we basically select pixels belonging to the galaxy (the object region) 
with the segmentation image, 
we add and exclude some pixels that belong to objects/substructures
detected in the ACS and NIRCam images across the boundary 
defined by the segmentation image.
For the adjustment, we also run SExtractor on the ACS and NIRCam images 
with a detection threshold of 1.3 times local background root mean square
over 25 connected pixels.
If more than 60\% of pixels for a source detected in the ACS and NIRCam 
data are included in the object region defined by the segmentation image,
we include all pixels of this source in the analysis and add some pixels
outside of the object region if exist.
On the other hand, if less than 40\% of pixels for the source are included,
we exclude all pixels of this source from the analysis as
another object.
We use the adjusted object region to measure the morphological indices
for each galaxy.

We estimate pixel-to-pixel background fluctuation in the 18'' $\times$ 18'' 
region of the ACS and NIRCam data with masking the object region and
pixels that belong to the other objects.
We then define pixels higher than 1$\sigma$ value of the background 
fluctuation in the object region as the object, 
and mask the other pixels less than 1$\sigma$ in the region.

We use the F814W, F115W, and F150W-band images for sample galaxies
at $0.2<z<0.7$, $0.7<z<1.3$, and $1.3<z<2.0$, respectively, 
in order to perform the morphological classification in the rest-frame
$V$ to $R$ band.
The morphological K-correction effects due to 
for example, colour differences between bulge and disk, blue star-forming
regions/clumps, dust extinction, and so on 
(e.g., \citealp{win02}; \citealp{mag18}) can be mitigated by analysing
galaxies at different redshifts in the similar rest-frame wavelengths.
Nevertheless, we note that relatively small effects such as the higher
asymmetry in the shorter rest-frame wavelengths within each redshift range
(i.e., rest-frame $V$ to $R$ band) could remain.

\begin{figure}
  \includegraphics[width=\columnwidth]{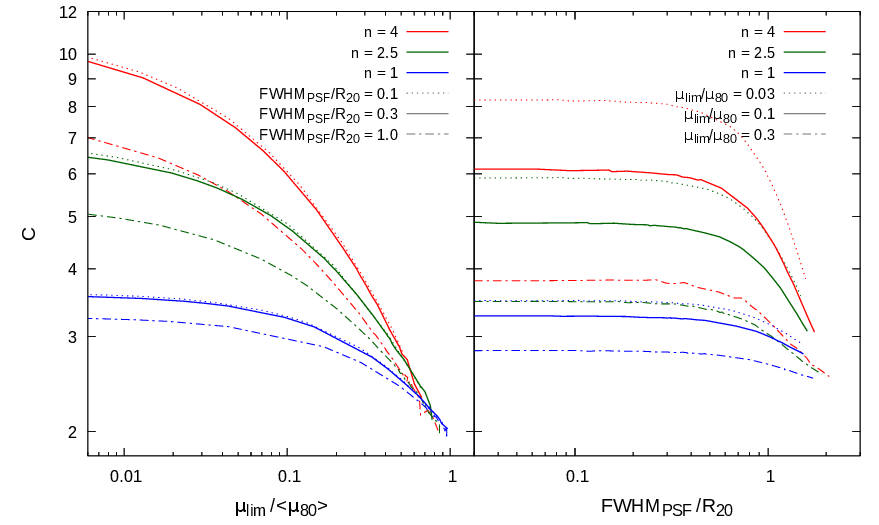}
  \caption{
    $C$ values for PSF-convolved S{\'e}rsic profiles
    with different S{\'e}rsic indices
     as a function of $\mu_{\rm lim}$/<$\mu_{80}$> (left) and
    FWHM$_{\rm PSF}$/$R_{20}$ (right).
    The red, green, and blue lines show those with $n=$ 1, 2.5, and 4,
    respectively. The dotted, solid, and dashed-dotted lines represent
    those for different FWHM$_{\rm PSF}$/$R_{20}$ (left) and
    $\mu_{\rm lim}$/<$\mu_{80}$> (right) values.
    \label{fig:modelC}}
\end{figure}

\subsubsection{Concentration}

In this study, we measure the concentration index defined as 
$C = R_{80}/R_{20}$, 
where $R_{80}$ and $R_{20}$ are radii which enclose 80\% and 20\% of
total flux of the object, in the same way as in \citet{him23}.
The total flux is estimated as a sum of the 
pixels higher than 1$\sigma$ value of the background fluctuation
in the object region.
Since the object region is defined by an extent of the object in
the $izYJHK_{s}$-band chi-square image, which is often more extended than
those in the ACS and NIRCam images, we correct for
contribution from background noise higher than the 1$\sigma$ value.
To estimate this background contribution, we make sky images by
replacing pixels in the object region with randomly selected ones
that do not belong to any objects in the 18'' $\times$ 18'' field.
All pixels outside the object region in the sky image are masked with
zero value.
We define pixels higher than 2$\sigma$ in the object image as
object-dominated region, and mask pixels at the same coordinates with
this region in the sky image, because no biased contribution from noises
is expected for those pixels with high object fluxes.
We sum up pixels higher than 1$\sigma$ value in the sky image
except for the object-dominated region, and subtract this from
the total flux of the object.

In order to estimate $R_{80}$ and $R_{20}$, we measure a growth curve
with elliptical apertures centred at a flux-weighted mean position of
the object pixels.
The shape of elliptical apertures is determined from second-order moments
around the centre of the all object pixels.
In the measurements of the growth curve, we perform   
the similar subtraction of the background contribution as in the total
flux measurements with the same sky images.
We make 50 sky images for each object and repeat the measurements of $C$,
and adopt their mean and standard deviation as $C$ and its uncertainty,
respectively.

\begin{figure}
  \includegraphics[width=1.05\columnwidth]{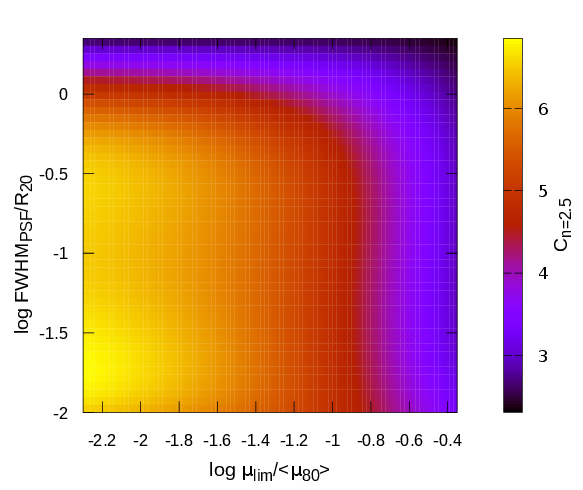}
  \caption{
    The fitted $C_{n=2.5}$, concentration index for the S{\'e}rsic profile with
    $n=2.5$, as a function of $\log(\mu_{\rm lim}/{\rm <}\mu_{80}{\rm >})$ and
    $\log({\rm FWHM}_{\rm PSF}/R_{20})$. 
  \label{fig:modelCn2p5}}
\end{figure}

The concentration index depends on both the surface brightness limit and
spatial resolution of the data (e.g., \citealp{kaj01}; \citealp{man10};
\citealp{wan24}). 
Since regions of the object where surface brightness is fainter than the 
observational limit could not be identified as those pixels belonging to
the object,   
the total fluxes tend to be underestimated for objects with lower surface
brightness,  
which leads to underestimates of mainly $R_{80}$, and therefore $C$.
On the other hand, if sizes of (inner part of) the objects are comparable
to the PSF size, their sizes, in particular $R_{20}$ could be significantly
overestimated, which also leads to underestimates of $C$.
In order to examine these effects, we calculate $C$ for idealised S{\'e}rsic
profiles with $n=$ 1, 2.5, and 4 as a function of
$\mu_{\rm lim}$/<$\mu_{80}$>
and FWHM$_{\rm PSF}$/$R_{20}$, where $\mu_{\rm lim}$ and <$\mu_{80}$> are
the surface brightness limit of the data and mean surface brightness within
$R_{80}$ of the object. 
We integrate the PSF-convolved S{\'e}rsic profiles down to $\mu_{\rm lim}$
to calculate the total flux. We then determine $R_{20}$ and $R_{80}$ from
the PSF-convolved profile and the total flux, 
and calculate $C$,  $\mu_{\rm lim}$/<$\mu_{80}$>, and
FWHM$_{\rm PSF}$/$R_{20}$ for these models.
In Figure \ref{fig:modelC}, one can see that $C$ systematically decreases
with increasing $\mu_{\rm lim}$/<$\mu_{80}$> and FWHM$_{\rm PSF}$/$R_{20}$,  
and $C$ values for different S{\'e}rsic indices become similar with each other
at high $\mu_{\rm lim}$/<$\mu_{80}$> and FWHM$_{\rm PSF}$/$R_{20}$ values.
Since $\mu_{\rm lim}$/<$\mu_{80}$> and FWHM$_{\rm PSF}$/$R_{20}$ can be directly
measured in the observed data, we can compare observed $C$ values with
these simulated $C$ values for the idealised S{\'e}rsic profiles.
In order to select galaxies with early-type morphology,
we use a criterion of $C > C_{n=2.5}$, where $C_{n=2.5}$ is a $C$ value of   
the PSF-convolved S{\'e}rsic profile with $n=2.5$ for given
$\mu_{\rm lim}$/<$\mu_{80}$> and FWHM$_{\rm PSF}$/$R_{20}$.
We calculate $C_{n=2.5}$ for various $\mu_{\rm lim}$/<$\mu_{80}$> and
FWHM$_{\rm PSF}$/$R_{20}$ values, and fit $C_{n=2.5}$ values with
a 3rd order polynomial function of $\mu_{\rm lim}$/<$\mu_{80}$> and
FWHM$_{\rm PSF}$/$R_{20}$ in logarithmic scale.
Figure \ref{fig:modelCn2p5} shows the fitted $C_{n=2.5}$ as a function of
$\mu_{\rm lim}$/<$\mu_{80}$> and FWHM$_{\rm PSF}$/$R_{20}$.
We calculate $C_{n=2.5}$ for each sample galaxy from 
its observed $\mu_{\rm lim}$/<$\mu_{80}$> and FWHM$_{\rm PSF}$/$R_{20}$ values,  
where FWHM$_{\rm PSF}$ and $\mu_{\rm lim}$ are those of the F814W-band,
F115W-band, and F150W-band data for sample galaxies 
at $0.2<z<0.7$, $0.7<z<1.3$, and $1.3<z<2.0$, respectively.
We then compare the observed $C$ value with $C_{n=2.5}$ for the galaxy,   
and select those galaxies with $C > C_{n=2.5}$ as candidates for
early-type galaxies.

\begin{figure}
  \includegraphics[width=1.0\columnwidth]{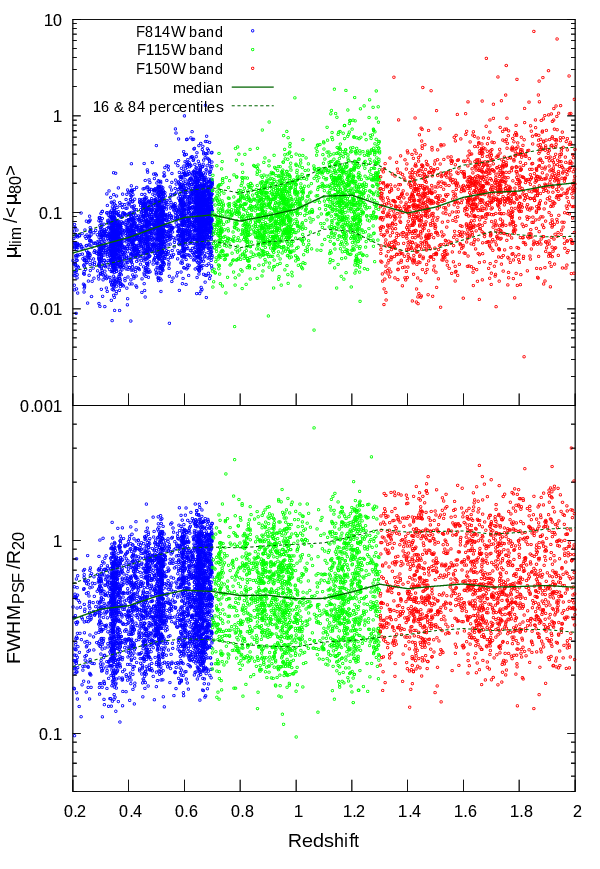}
  \caption{
    FWHM$_{\rm PSF}$/$R_{20}$ and $\mu_{\rm lim}$/<$\mu_{80}$> as a function
    of redshift for sample galaxies with $M_{\rm star} > 10^{10.5} M_{\odot}$
    at $0.2<z<2.0$.
    The F814W, F115W, and F150W-band data are used for 
    those galaxies at $0.2<z<0.7$, $0.7<z<1.3$, and $1.3<z<2.0$
    (red, green, and blue dots), respectively.
    The solid line shows the median value in each redshift bin with a width
    of $\pm$ 0.1, while the dashed lines represent the 16 \& 84 percentiles.  
  \label{fig:SB_PSFR20}}
\end{figure}

Figure \ref{fig:SB_PSFR20} shows 
$\mu_{\rm lim}$/<$\mu_{80}$> and FWHM$_{\rm PSF}$/$R_{20}$ 
distributions as a function of redshift for sample galaxies.
While most galaxies have $\mu_{\rm lim}$/<$\mu_{80}$> $\lesssim 0.5$ and
FWHM$_{\rm PSF}$/$R_{20}$ $\lesssim 1$,
there are a small fraction of galaxies with $\mu_{\rm lim}$/<$\mu_{80}$>
$\gtrsim 1$ or FWHM$_{\rm PSF}$/$R_{20}$ $\gtrsim 1.5$, especially,
at $z \gtrsim 1.5$.
The classification with $C$ could be unreliable for such objects with  
low-surface brightness or small size, because the differences
in $C$ become very small (Figure \ref{fig:modelC}).
In order to consider the uncertainty in the morphological classification, 
we calculate lower and upper limits of the early-type fraction by 
counting numbers of galaxies with $C-\Delta C > C_{n=2.5}$ and those
with $C+\Delta C > C_{n=2.5}$, respectively,
where $\Delta C$ is the uncertainty of $C$ mentioned above.
Those galaxies with $\mu_{\rm lim}$/<$\mu_{80}$>
$\gtrsim 1$ or FWHM$_{\rm PSF}$/$R_{20}$ $\gtrsim 1.5$
tend to show similar $C$ values irrespective of their intrinsic surface
brightness profiles, and therefore they tend to have $C+\Delta C > C_{n=2.5}$,
but not to satisfy  $C-\Delta C > C_{n=2.5}$.

\begin{figure}
  \includegraphics[width=\columnwidth]{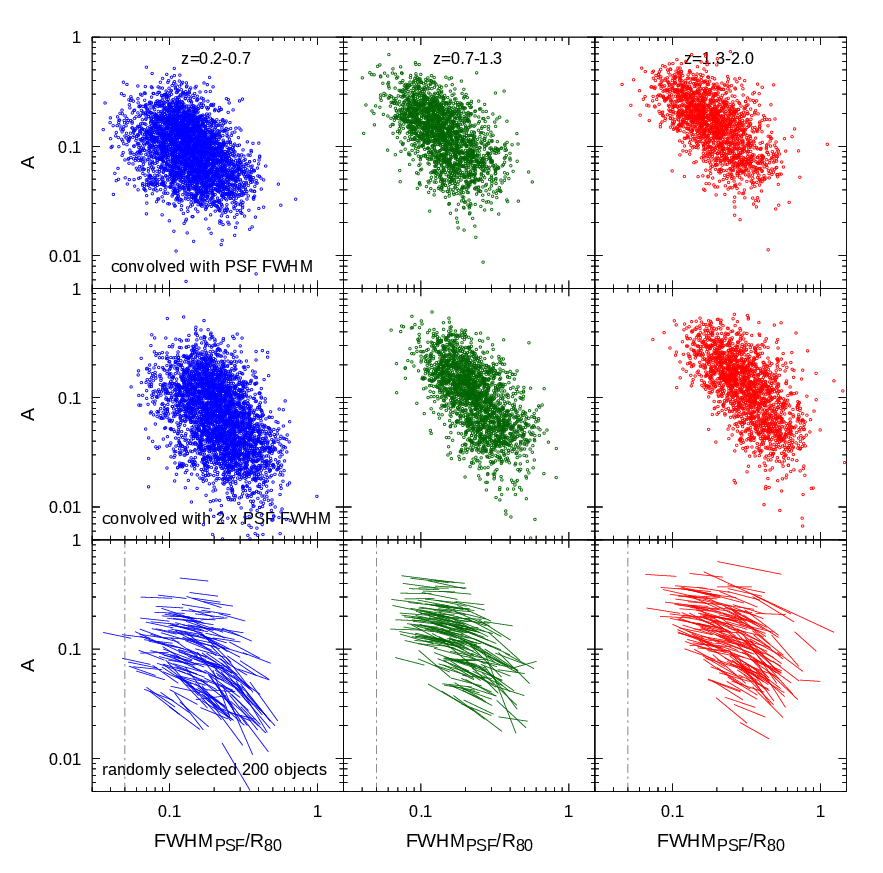}
  \caption{
    FWHM$_{\rm PSF}/R_{80}$ vs. $A$ diagrams for sample galaxies in the
    {\it HST} or {\it JWST} images convolved with
    the Gaussian kernels whose FWHMs are the same as PSF (top) and two times 
    larger than PSF (middle). The bottom panels show the differences of 
    FWHM$_{\rm PSF}/R_{80}$ and $A$ between those images convolved with
    the different kernels for 200 objects randomly selected in each redshift
    bin. Colours of the symbols represent the F814W (blue), F115W (green),
    and F150W-band (red) data.
  \label{fig:simA}}
\end{figure}

\begin{figure}
  \includegraphics[width=\columnwidth]{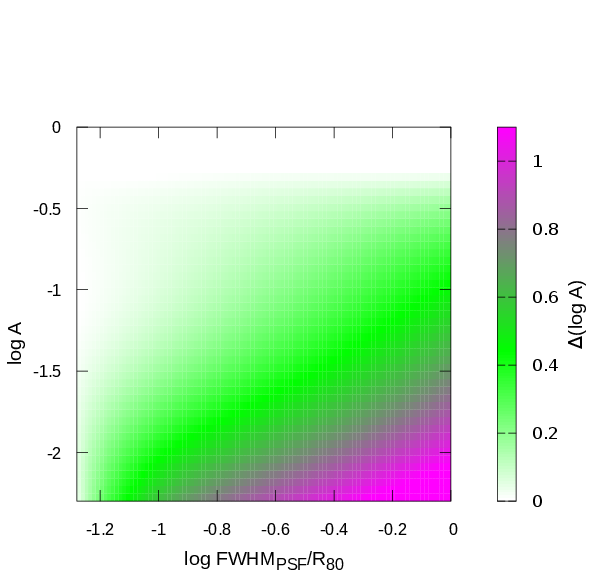}
  \caption{
    Correction for the asymmetry index, $\Delta(\log{A})$ as a function of
    $\log{{\rm FWHM}_{\rm PSF}/R_{80}}$ and $\log{A}$.  
  \label{fig:modeldA}}
\end{figure}

\subsubsection{Asymmetry}

In addition to $C$, we use the asymmetry index $A$ to select galaxies
with early-type morphology.
We measure $A$ for sample galaxies with the same method as in \cite{him23}.
We adopt the same definition of the object pixels
as in the calculation of $C$, namely, 
those pixels higher than 1$\sigma$ value in the object region.
We rotate the image by 180 degree and subtract it from the
original image. 
We then sum up only pixels with positive values in the subtracted image, 
and calculate fraction of this total positive flux in the 
rotation-subtracted image relative to the total flux of the object in the 
original image as $A$.
Following \citet{con00}, 
we search a rotation centre for which the positive total flux
in the rotation-subtracted image is minimum with a step of 0.5 pixel 
in X and Y axes, and adopt it.
With the same sky images for the object described in the previous subsection, 
we estimate contribution from the background noise
in the rotation-subtracted image, and correct the positive total flux
for it as in \cite{him23}.
In the calculation of the noise contribution, we exclude similar 
  (asymmetric) object-dominated regions where pixel values
  in the rotation-subtracted object image are higher than 2$\sigma$ 
  to avoid the overestimate of the noise contribution.
We use the same total flux of the object as in the calculation of $C$,    
which is also corrected for the contribution from the background noise.
By using the 50 random sky images, 
we repeat the calculations described above and
adopt the mean and standard deviation as $A$ and its uncertainty,
respectively.

\begin{figure*}
  \includegraphics[width=2.0\columnwidth]{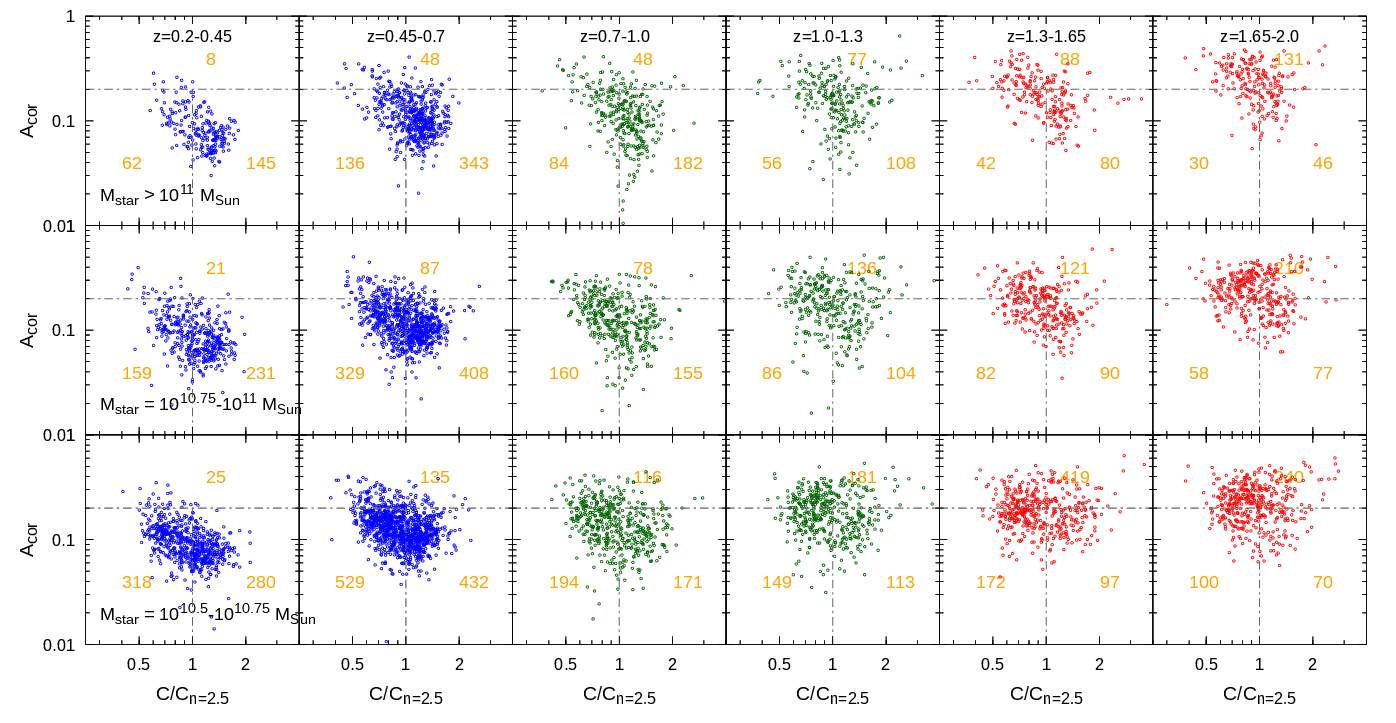}
  \caption{
    $C/C_{n=2.5}$ vs. $A_{\rm cor}$ diagram for sample galaxies in the different
    stellar mass and redshift bins.
    The dashed-dotted lines show the boundaries for the morphological
    classification of $C =  C_{n=2.5}$ and $A_{\rm cor} = 0.2$.
    Colours of the symbols represent the F814W (blue), F115W (green),
    and F150W-band (red) data.
    The numbers of galaxies with the three morphological types are also
      shown in each panel.
  \label{fig:selearly}}
\end{figure*}

The asymmetry index also depends on the spatial resolution of the data
(e.g., \citealp{con00}; \citealp{lot04}; \citealp{tho21};
  \citealp{saz24}; \citealp{luo25}).
In order to examine the resolution effects on $A$, we convolve the F814W,
F115W, and F150W-band images with two Gaussian kernels, namely,
those with the PSF FWHM and 2 $\times$ PSF FWHM, and then measure $A$
in the both convolved images with the same manner as in the original images.
The top and middle panels of Figure \ref{fig:simA} show $A$ measured in
the images convolved with the kernels with the PSF FWHM and 2 $\times$
PSF FWHM, respectively, as a function of FWHM$_{\rm PSF}$/$R_{80}$.
In the images convolved with the kernel with the larger FWHM, the $A$ values 
tend to be lower and FWHM$_{\rm PSF}$/$R_{80}$ is larger.
The bottom panel shows differences of $A$ measured in the images convolved
with the two kernels for randomly selected 200 objects in each redshift
range. Note that we use the images convolved with PSF as the sharper image
to mitigate the effects of pixel-to-pixel noise.
In all the redshift ranges, the decreases in $A$ tend to be larger
in smaller galaxies with a relatively high FWHM$_{\rm PSF}$/$R_{80}$, 
which reflects that the asymmetric features of such small galaxies are
more easily smeared out by PSF.
In order to correct the resolution effect empirically, we calculate a slope, 
namely, $d(\log{A})/d(\log{{\rm FWHM}_{\rm PSF}/R_{80}})$ for all sample
galaxies, and fit the slopes with a 3rd order polynomial of $\log{A}$ and
$\log{{\rm FWHM}_{\rm PSF}/R_{80}}$ measured in the images convolved with
the Gaussian kernel with 2 $\times$ PSF FWHM. By using the calculated
$d(\log{A})/d(\log{{\rm FWHM}_{\rm PSF}/R_{80}})$ as a function of
$\log{A}$ and $\log{{\rm FWHM}_{\rm PSF}/R_{80}}$, we integrate 
$\Delta(\log{A})$ from the observed values of $\log{A}$ and
$\log{{\rm FWHM}_{\rm PSF}/R_{80}}$
to $\log{{\rm FWHM}_{\rm PSF}/R_{80}} = -1.3$ for each object.
Figure \ref{fig:modeldA} shows the estimated $\Delta(\log{A})$
as a function of $\log{A}$ and $\log{{\rm FWHM}_{\rm PSF}/R_{80}}$.
The correction is larger for objects with larger FWHM$_{\rm PSF}$/$R_{80}$
(i.e., smaller size relative to PSF) and lower $A$.
We correct $A$ values by $\log{A_{\rm cor}} = \log{A} + \Delta(\log{A})$,
and used $A_{\rm cor}$ for morphological classification.

We note that the PSF FWHMs of the F814W, F115W, and F150W-band data
  correspond to 0.32--0.69, 0.47-0.55, and 0.58--0.59 kpc for galaxies
  at $0.2<z<0.7$, $0.7<z<1.3$, and $1.3<z<2.0$, respectively.
  The spatial resolution of the data are slightly different in physical
  scale among the sample galaxies at different redshifts.
  On the other hand, the bottom panels of Figure \ref{fig:simA} show
  that the resolution effect on $A$ is relatively small at
  $\log{{\rm FWHM}_{\rm PSF}/R_{80}} \lesssim -1$ irrespective of redshift.
  Since we correct $A$ for the resolution effect so that $A_{\rm cor}$
  become a value at $\log{{\rm FWHM}_{\rm PSF}/R_{80}} = -1.3$, 
  we expect that the slight differences of the spatial resolution
  in physical scale do not significantly affect on $A_{\rm cor}$.

  Several previous studies reported that $A$ tends to be underestimated
    when a S/N ratio of the image is relatively low
    (e.g., \citealp{tho21}; \citealp{saz24}).
    While different definitions of the asymmetry index that is more immune
    to the noise effects, for example, Root Mean Square asymmetry $A_{\rm RMS}$
    (\citealp{deg23}; \citealp{saz24})
    have been proposed, we use the original definition of $A$ corrected
    for the resolution effect in this study, 
    because it simply reflects flux contribution of the asymmetric features.
    Although we carefully correct the contribution from the background noise
    as described above, we need to keep in mind that 
    the $A_{\rm cor}$ values of sample galaxies
    with relatively low surface brightness could be underestimated.

We use a criterion of $A_{\rm cor}<0.2$ to select early-type morphology, 
since most of quiescent galaxies at $z\sim 0.8$ show $A<0.2$ in \cite{him23},
where $A$ is measured with the similar manner.
Since the criterion of $A_{\rm cor} < 0.2$ is somehow arbitrary,
we examine cases with $A_{\rm cor}<0.15$ and $A_{\rm cor}<0.25$
to check effects of the slightly
different values of the criterion.
We also count the numbers of sample galaxies with $C-\Delta C > C_{n=2.5}$
and $A+\Delta A_{\rm cor} < 0.2$ and those with $C+\Delta C > C_{n=2.5}$ and
$A_{\rm cor}-\Delta A_{\rm cor} < 0.2$, where $\Delta A_{\rm cor}$ is
the uncertainty of $A_{\rm cor}$, 
to calculate the lower and upper limits of the
early-type fraction taking the uncertainty of $A$ into account.

\begin{figure}
  \includegraphics[width=\columnwidth]{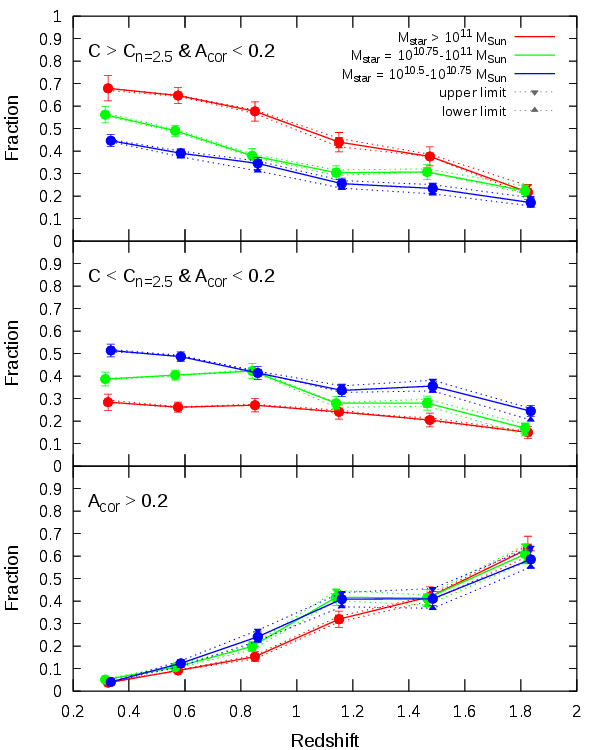}
  \caption{
    {\bf top:} Fraction of early-type galaxies with $C > C_{n=2.5}$ and
    $A_{\rm cor} < 0.2$
    in sample galaxies with $M_{\rm star} > 10^{10.5} M_{\odot}$ as a function of
    redshift.
    The red, green, and blue circles and solid lines
    show the early-type fraction
    for galaxies with $M_{\rm star} > 10^{11} M_{\odot}$,
    $M_{\rm star} = 10^{10.75}$--$10^{11} M_{\odot}$, and
    $M_{\rm star} = 10^{10.5}$--$10^{10.75} M_{\odot}$, respectively.
    The error bars are based on the Poisson statistics.
    The triangles and dotted lines represent the lower and upper limits
    of the fraction, which are estimated from the numbers of those galaxies
    with $C-\Delta C > C_{n=2.5}$ and $A_{\rm cor} + \Delta A_{\rm cor} < 0.2$ and
    those with $C+\Delta C > C_{n=2.5}$ and $A_{\rm cor} - \Delta A_{\rm cor} < 0.2$,  respectively.
    {\bf middle:} Fraction of late-type (disk-dominated) galaxies with
    $C < C_{n=2.5}$ and $A_{\rm cor} < 0.2$ as a function of redshift.
    The symbols are the same as the top panel.
    The triangles and dotted lines represent the lower and upper limits
    estimated from the numbers of those galaxies with $C+\Delta C < C_{n=2.5}$
    and $A_{\rm cor} + \Delta A_{\rm cor} < 0.2$ and those with
    $C-\Delta C < C_{n=2.5}$ and
    $A_{\rm cor} - \Delta A_{\rm cor} < 0.2$.
    {\bf bottom:} Fraction of irregular galaxies with $A_{\rm cor} > 0.2$.
    The symbols are the same as the top panel.
    The triangles and dotted lines represent the lower and upper limits
    estimated from the numbers of those galaxies with
    $A_{\rm cor} - \Delta A_{\rm cor} > 0.2$ and those with
    $A_{\rm cor} + \Delta A_{\rm cor} > 0.2$.
  \label{fig:frac_morph}}
\end{figure}
 
\subsubsection {Minimum $R_{20}$ and Concentration of asymmetric features}
 
We also use minimum $R_{20}$ and 
central concentration of asymmetric features, $C_{A}$
in the following analyses, 
while these indices are not used to select early-type galaxies.
We search minimum $R_{20}$ by estimating $R_{20}$, 
within which 20\% of the total flux of the object is included,
with varying the centre position of the aperture, while the aperture centre 
is fixed at the flux-weighted mean position in the calculation of $C$.
$R_{20}$ basically represents the surface brightness or stellar mass
density in the central region of the galaxy, but the flux-weighted mean
position does not necessarily correspond to the brightest/highest-density
region for irregular/peculiar galaxies.
The minimum $R_{20}$ is expected to be more closely related 
with the surface brightness/stellar mass density in the brightest/densest
region even for such galaxies with significant asymmetric features.

$C_{A}$ is a combination of the asymmetry and concentration indices,
i.e., a concentration index measured on the rotation-subtracted images
described above, and 
\cite{him23} devised it to differentiate asymmetric features such as central
disturbances, tidal tails, star-forming regions on a normal disk, and so on.
We similarly estimate radii which contain 80\% and 20\% of the total
positive flux in the rotation-subtracted object image,
namely, $R_{A,80}$ and $R_{A,20}$, 
and calculate $C_{A} = R_{A,80}/R_{A,20}$.
We use the same rotation-subtracted images as in the calculation of $A$,
for which the rotation centre is determined to obtain the minimum value of $A$.
We measure a growth curve on those images with circular apertures centred
at the rotation centre.
In measurements of $r_{A,80}$ and $r_{A,20}$, we also correct for  
the contribution from the background noise 
with the same rotation-subtracted sky image used in the previous subsection.
We repeat the calculations with the 50 random sky images and
adopt the mean and standard deviation as $C_{A}$ and its uncertainty,
respectively.

\begin{figure}
  \includegraphics[width=\columnwidth]{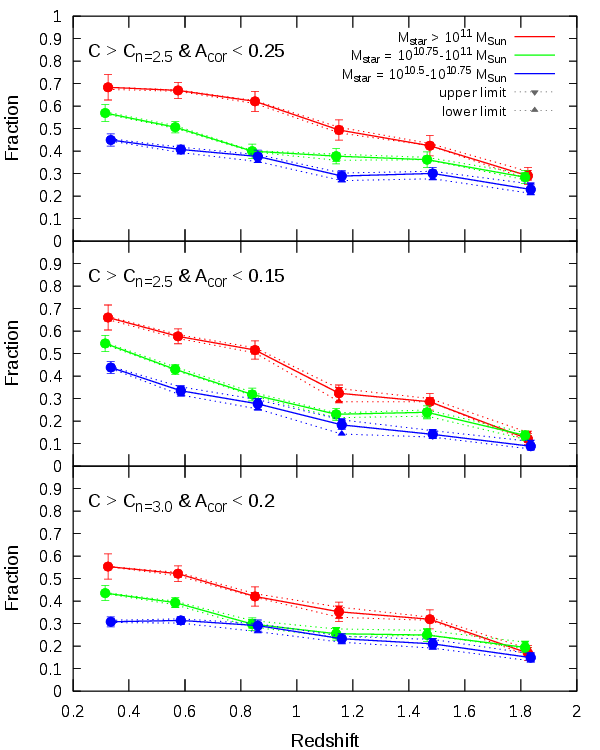}
  \caption{
    Evolution of the fraction of early-type galaxies selected with 
    slightly different criteria, namely, 
    $C > C_{n=2.5}$ and $A_{\rm cor} < 0.25$ (top), $C > C_{n=2.5}$
    and $A_{\rm cor} < 0.15$ (middle), 
    and $C > C_{n=3.0}$ and $A_{\rm cor} < 0.2$ (bottom).
    The symbols are the same as Figure \ref{fig:frac_morph}.
  \label{fig:n3_diffA}}
\end{figure}

\section{Results} \label{sec:res}

\subsection{Fraction of Early-type Galaxies}

Figure \ref{fig:selearly} shows $C/C_{n=2.5}$ vs. $A_{\rm cor}$ diagram for our
sample galaxies with $M_{\rm star} > 10^{10.5} M_{\odot}$ at $0.2<z<2.0$.
We measure $C$ and $A$ indices in the wavelengths that correspond to
the rest-frame $V$ to $R$ band for those galaxies as described in the
previous section.
One can see that those galaxies tend to have lower $C$ and higher $A$
values at higher redshifts.
We use the criteria of $C > C_{n=2.5}$ and $A_{\rm cor} < 0.2$
(i.e., bottom-right region in the panels) 
to select galaxies with early-type morphology.

The top panel of Figure \ref{fig:frac_morph} 
shows fraction of those galaxies with $C > C_{n=2.5}$ and $A_{\rm cor} < 0.2$
in our sample galaxies with different stellar masses
as a function of redshift.
We estimate lower and upper limits of the fraction by counting
the numbers of those galaxies with $C-\Delta C > C_{n=2.5}$ 
and $A_{\rm cor} + \Delta A_{\rm cor} < 0.2$ and those with
$C+\Delta C > C_{n=2.5}$ and
$A_{\rm cor} - \Delta A_{\rm cor} < 0.2$,
which are shown as the dotted lines in the panel.
The widths between the lower and upper limits are smaller than
the Poisson errors of the fraction in most stellar mass and redshift bins,
which means that the uncertainty in the morphological classification does not
strongly affect the result.

In the top panel of the figure,
the fraction of those galaxies with $C > C_{n=2.5}$ and $A_{\rm cor} < 0.2$
decreases with increasing redshift in all the stellar mass bins.
The early-type fraction is the highest in the most 
massive galaxies with $M_{\rm star} > 10^{11} M_{\odot}$ at least at $z < 1.65$.
The early-type fraction in those massive galaxies decreases with increasing 
redshift from
$\sim$ 70\% at $z \sim 0.3$ to 20--25\% at $z \sim 1.8$.
Less massive galaxies show a similar evolution of the fraction but at 
lower values of the fraction.
The fraction in galaxies with $M_{\rm star} = $ $10^{10.75}$--$10^{11} M_{\odot}$
($10^{10.5}$--$10^{10.75} M_{\odot}$) decreases with increasing redshift from
50--60\% (40--50\%) at $z \sim 0.3$ to 20--25\% (15--20\%) at
$z \sim 1.8$.

\begin{figure}
  \includegraphics[width=0.97\columnwidth]{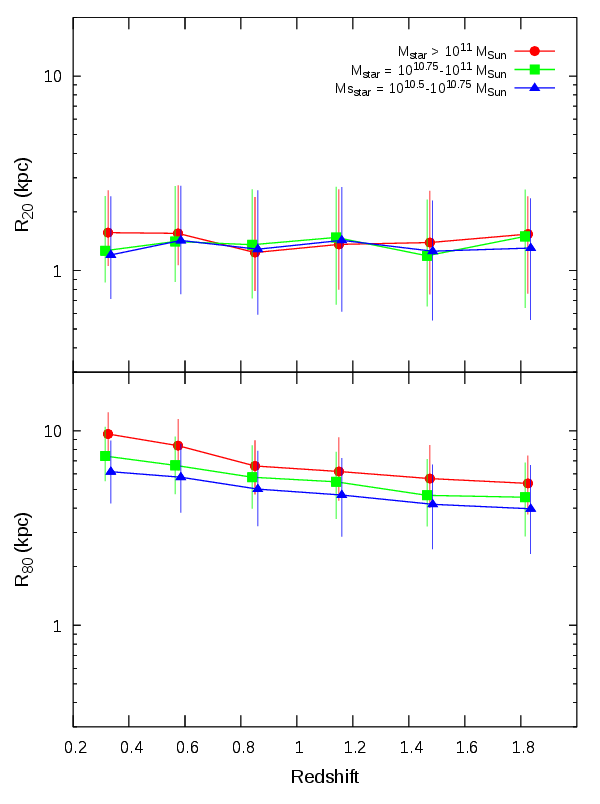}
  \caption{
    Evolution of $R_{20}$ (top) and $R_{80}$ (bottom),  
    semi-major radii which enclose 20\% and 80\% of the total flux of
    the object,   
    for sample galaxies in the different stellar mass bins.
    The circles, squares, and triangles represent median values of $R_{20}$ 
	or $R_{80}$ for those galaxies with $M_{\rm star} > 10^{11} M_{\odot}$, 
	$M_{\rm star} = 10^{10.75}$--$10^{11} M_{\odot}$, and 
	$M_{\rm star} = 10^{10.5}$--$10^{10.75} M_{\odot}$, respectively.
    The error bars show 16 and 84 percentiles of $R_{20}$ or $R_{80}$ in 
	the redshift bins.
  \label{fig:r2080_z}}
\end{figure}

We also examine the fraction of galaxies that do not satisfy the criteria of
$C > C_{n=2.5}$ and $A_{\rm cor} < 0.2$ as a function of redshift.
The middle and bottom panels show 
the fractions of galaxies with
$C < C_{n=2.5}$ and $A_{\rm cor} < 0.2$, and those with $A_{\rm cor} > 0.2$,
respectively.
The fraction of late-type (disk-dominated) galaxies
with $C < C_{n=2.5}$ and $A_{\rm cor} < 0.2$ decreases with increasing stellar
mass in all redshift bins.
The fraction of these late-type galaxies seems to decrease
with increasing redshift, 
especially at $z \gtrsim 1$, but the evolution is weaker than that of early-type
galaxies.
In contrast, the fraction of irregular galaxies with $A_{\rm cor} > 0.2$ clearly
increases with increasing redshift from $\sim$ 5\% at $z \sim 0.3$ to
$\sim$ 60--65\% at $z \sim 1.8$.
The fraction of irregular galaxies shows a weaker stellar mass dependence than
early-type galaxies and disk-dominated ones.
The fraction of irregular galaxies seems to be independent of stellar mass
or slightly higher in more massive galaxies 
at $z>1.3$, while the fraction tends to be lower in more massive galaxies
at $z<1.3$.
Relatively weak evolution of the fraction of irregular galaxies
at $z\sim0.7$ and $z\sim1.3$ could be caused by the morphological
K-correction, because $A$ is measured in slightly longer rest-frame wavelengths
 for galaxies at $0.7<z<1.0$ and $1.3<z<1.65$ than those galaxies at
$0.45<z<0.7$ and $1.0<z<1.3$.

 Figure \ref{fig:n3_diffA} shows the evolution of the early-type fraction
 estimated with slightly different criteria of $C$ and $A$.
 In the top and middle panels of the figure, we show the cases with
 $A_{\rm cor} < 0.25$ and $A_{\rm cor} < 0.15$, respectively, rather than
 $A_{\rm cor} < 0.2$.
Even if we adopt a slightly different criterion of $A$, 
the systematic decrease of the early-type fraction with increasing redshift
is not strongly affected.
Since the fraction of those with $C > C_{n=2.5}$ and $A_{\rm cor} > 0.2$ is
relatively small over the redshift range we investigated (Figure
\ref{fig:selearly}), 
a looser criterion of $A_{\rm cor} < 0.25$ leads to a small increase of
the early-type fraction at higher redshifts, which could make the
redshift evolution sightly weaker.
On the other hand, a more stringent criterion of $A_{\rm cor} < 0.15$ slightly
decreases the early-type fraction preferentially at higher redshifts, 
which could slightly strengthen the evolution. 
We also examine a slightly different criterion for $C$, namely
$C > C_{n=3.0}$, where $C_{n=3.0}$ is $C$ value expected for the S{\'e}rsic 
profile with $n=3.0$ rather than $n=2.5$.
In the bottom panel, the early-type fraction is slightly lower than
the case with $C> C_{n=2.5}$ irrespective of stellar mass and redshift,
but the redshift evolution and stellar mass dependence of the fraction
do not strongly change.

\begin{figure}
  \includegraphics[width=1.05\columnwidth]{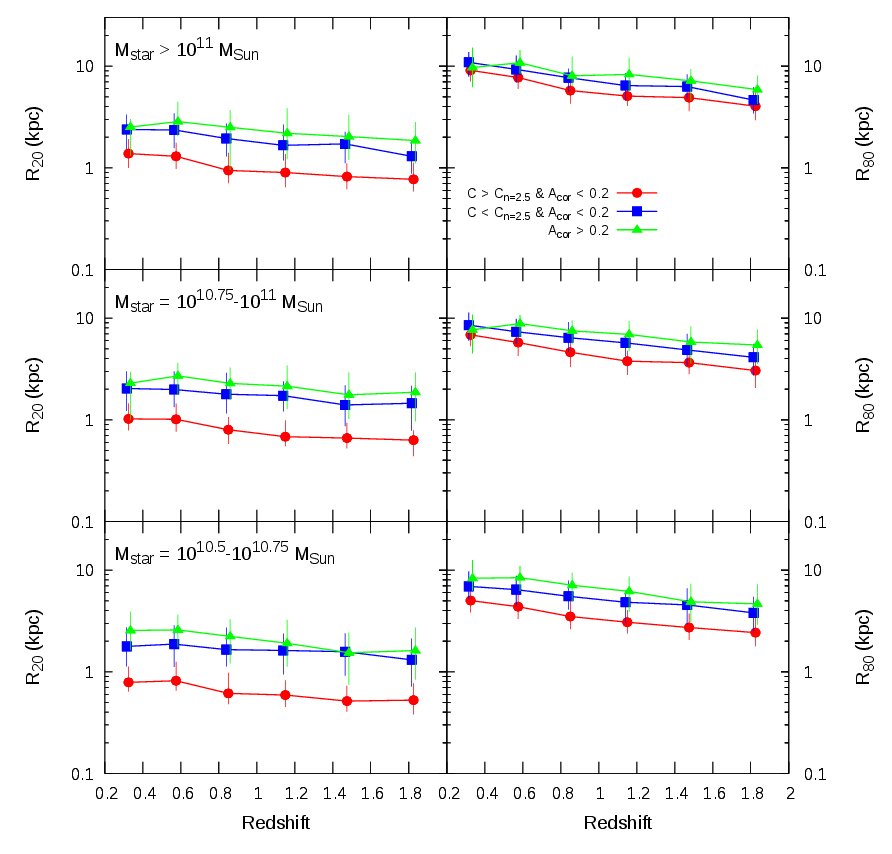}
  \caption{
    Evolution of $R_{20}$ (left) and $R_{80}$ (right)  
    for sample galaxies with the different morphological types
    and stellar masses.
    The circles, squares, and triangles represent median values of $R_{20}$
    or $R_{80}$ for early-type galaxies with $C > C_{n=2.5}$ and
    $A_{\rm cor} < 0.2$,  
    late-type galaxies with $C < C_{n=2.5}$ and $A_{\rm cor} < 0.2$, and irregular
    galaxies with $A_{\rm cor} > 0.2$, respectively.
    The error bars show 16 and 84 percentiles of $R_{20}$ or $R_{80}$ in
    the redshift bins.
    Stellar mass decreases from the top panels to the bottom panels.
  \label{fig:r2080_morph}}
\end{figure}

\subsection{Evolution and Mass Dependence of $R_{20}$ and $R_{80}$}

We here focus on evolution of $R_{80}$ and $R_{20}$ for our sample galaxies
in order to investigate how the fraction of early-type galaxies with high
concentration in their surface brightness increases with time.
Figure \ref{fig:r2080_z} shows $R_{20}$ and $R_{80}$ as a function of redshift
for sample galaxies in the different stellar mass bins.
One can see that $R_{80}$ shows a significant redshift evolution and
stellar mass dependence.
The median values of $R_{80}$ for those galaxies in the three stellar mass
bins, namely, $M_{\rm star} > 10^{11} M_{\odot}$,
$M_{\rm star} = $ $10^{10.75}$--$10^{11} M_{\odot}$, and
$M_{\rm star} = $ $10^{10.5}$--$10^{10.75} M_{\odot}$, increase with time 
from $z \sim 1.8$ to $z \sim 0.3$ by 0.25, 0.21, and 0.19 dex, respectively.
The median $R_{80}$ also increases with increasing stellar mass by
$\sim $ 0.13--0.19 dex between $M_{\rm star} > 10^{11} M_{\odot}$ and
$M_{\rm star} = 10^{10.5}$--$10^{10.75} M_{\odot}$ depending on redshift.
On the other hand, the median values of $R_{20}$ in all stellar mass bins
are 1.2--1.5 kpc in the same redshift range,
and show no significant redshift evolution and stellar mass dependence.
A width between 16 and 84 percentiles of $R_{20}$ is $\sim 0.4$--0.6 dex, 
which is systematically larger than that of $R_{80}$ ($\sim$ 0.3--0.4 dex).
The widths of $R_{20}$ and $R_{80}$ seem to increase with 
increasing redshift by $\sim$ 0.1 dex between $z\sim0.3$ and
$z\sim1.8$, while they also increase with decreasing stellar mass
by $\sim$ 0.1 dex between $M_{\rm star} > 10^{11} M_{\odot}$ and
$M_{\rm star} = 10^{10.5}$--$10^{10.75} M_{\odot}$.

Figure \ref{fig:r2080_morph} shows the evolution of $R_{20}$ and $R_{80}$ for 
sample galaxies with the different morphological types separately.
While early-type galaxies tend to have smaller sizes than late-type and
irregular galaxies in both $R_{20}$ and $R_{80}$,
the variations among the three morphological types are systematically
larger in $R_{20}$.
The median $R_{20}$ of early-type galaxies is smaller than late-type galaxies
(irregular galaxies) by $\sim$ 0.25--0.45 dex ($\sim$ 0.3--0.6 dex) depending
on stellar mass and redshift.
These differences in median $R_{20}$ among the different types are comparable
to or larger than a width between 16 and 84 percentiles of $R_{20}$ in
early-type galaxies ($\sim$ 0.25--0.35 dex).
On the other hand, differences in median $R_{80}$ between
early-type galaxies and late-type (irregular) ones are $\sim$ 0.06--0.2 dex
($\sim$ 0.1--0.35 dex), which are similar with or smaller than a width
of $R_{80}$ distribution in early-type galaxies ($\sim$ 0.2--0.3 dex).
The differences in $R_{20}$ and $R_{80}$ among the different morphological types
 slightly decrease with increasing stellar mass.
In each morphological type, the redshift evolution in $R_{80}$ is
stronger than that in $R_{20}$.
In a fixed stellar mass range, the strength of the evolution is similar
among the three morphological types, but the size evolution of early-type
galaxies is slightly stronger at $z\lesssim 1$.

While $R_{80}$ shows the significant redshift evolution and
mass dependence, which could be related with the evolution and mass
dependence of the early-type fraction, 
the variations among galaxies with the different morphological types
are more clearly seen in $R_{20}$.

\begin{figure}
  \includegraphics[width=\columnwidth]{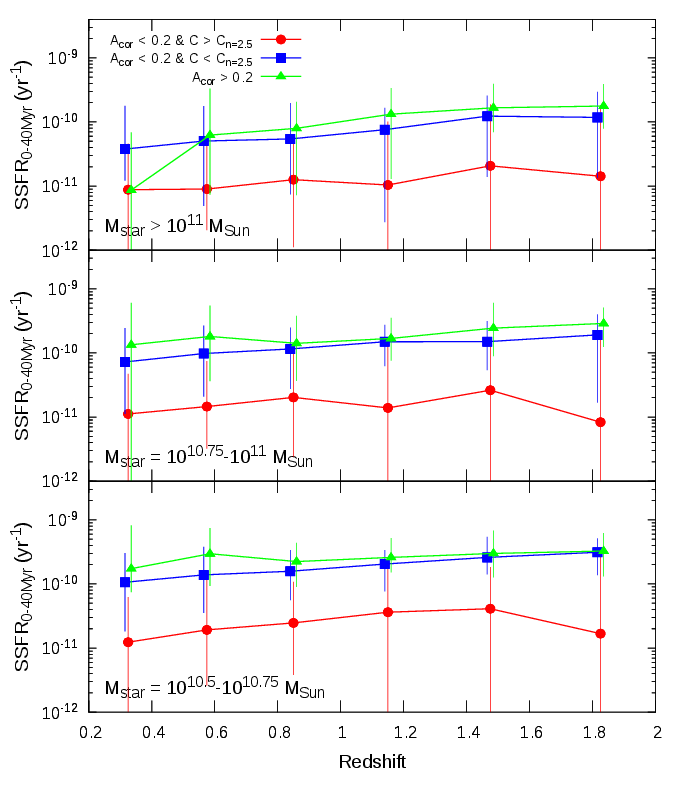}
  \caption{
    SSFR$_{\rm 0-40Myr}$ as a function of redshift for sample galaxies
    with the different morphological types and stellar masses.
    The symbols are the same as Figure \ref{fig:r2080_morph}.
  \label{fig:ssfr_morph}}
\end{figure}

\begin{figure}
  \includegraphics[width=\columnwidth]{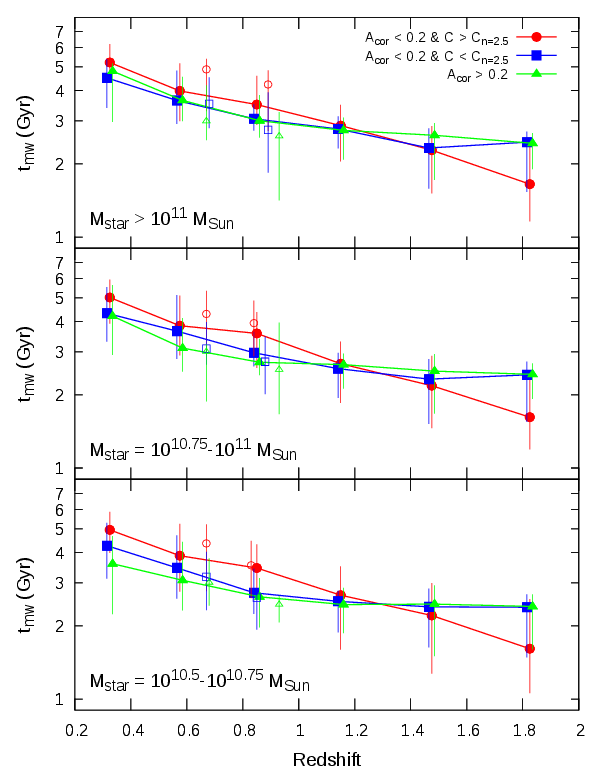}
  \caption{
    Mass-weighted mean stellar age $t_{\rm mw}$ as a function of redshift for
    sample galaxies
    with the different morphological types and stellar masses.
    The solid symbols are the same as Figure \ref{fig:r2080_morph}.
      The open circles, squares, and triangles represent $t_{\rm mw}$
      estimated from both the photometric SED and optical spectrum from
      the LEGA-C survey for early-type, late-type, and irregular galaxies,
      respectively. These open symbols are plotted at a median redshift
      of galaxies in each morphological and redshift bin.
    \label{fig:mwage_morph}}
\end{figure}

\subsection{SSFR and mean stellar age} \label{sec:ssfrage}

In order to investigate relationship between stellar mass assembly and the
formation of early-type morphology, we here examine SSFR and mean
stellar age as a function of redshift for sample galaxies with the different
morphological types.
Figure \ref{fig:ssfr_morph} shows SSFR$_{\rm 0-40Myr}$ as a function of
redshift for sample galaxies with the different morphological types and
stellar masses.
One can see that early-type galaxies have systematically lower SSFRs
than late-type and irregular galaxies in all stellar mass and redshift bins.
While late-type and irregular galaxies tend to be star-forming galaxies
with median SSFR$_{\rm 0-40Myr} \gtrsim 10^{-10}$ yr$^{-1}$,
the median values of SSFR$_{\rm 0-40Myr}$ for early-type galaxies are
$\sim 10^{-11}$--$10^{-10.5}$ yr$^{-1}$ and $\sim 1$ dex lower than
those of late-type and irregular galaxies.
Irregular galaxies tend to have slightly higher SSFRs than late-type
(disk-dominated) galaxies, except for the most massive and lowest redshift 
bin.
The number of massive irregular galaxies with
$M_{\rm star} > 10^{11} M_{\odot}$ at $0.2<z<0.45$ is eight  
(Figure \ref{fig:selearly}), and therefore their median SSFR
could have large uncertainty. Most of them have relatively low SSFRs
of $< 10^{-10.5} yr^{-1}$ and morphologies with double components,
which suggests they are dry mergers.
The median SSFRs gradually increase with increasing redshift for 
all the morphological types and stellar mass bins, although the evolution
  of early-type galaxies is not significant due to
  their relatively large errors at low SSFRs.
More massive galaxies tend to have slightly lower SSFRs, but the
strength of the redshift evolution is similar among galaxies with
different stellar masses. 

\begin{table}
	\centering
	\caption{The numbers of sample galaxies with the $HST$/$JWST$ data and LEGA-C spectra in the stellar mass and redshift bins.}
	\label{tab:spec}
	\begin{tabular}{cccc} 
		\hline
		stellar mass & type  & $z=$ 0.5--0.7 & $z=$ 0.7--1.0 \\
		\hline
		$M_{\rm star} > 10^{11} M_{\odot}$ & early-type & 115 & 74\\
                & late-type & 47 & 39\\
                & irregular & 21 & 25\\
                \hline
		$M_{\rm star} = 10^{10.75}$--$10^{11} M_{\odot}$ & early-type & 93 & 26\\
                & late-type & 67 & 46\\
                & irregular & 20 & 21\\
                \hline
		$M_{\rm star} = 10^{10.5}$--$10^{10.75} M_{\odot}$ & early-type & 60 & 14\\
                & late-type & 90 & 32\\
                & irregular & 24 & 16\\
		\hline
	\end{tabular}
\end{table}

Figure \ref{fig:mwage_morph} shows mass-weighted mean stellar age
as a function of redshift for galaxies with the different morphological types. 
While differences in mean stellar age among the different morphological types
are smaller than those in SSFR, early-type galaxies show stronger evolution
than late-type and irregular galaxies in all the stellar mass bins.
The weaker age evolution of late-type and irregular galaxies is consistent
with their higher SSFRs, because continuous star formation adds young
stellar population to these galaxies, which can weaken the evolution of
mean age from that expected for passive evolution.
At $z \lesssim 1$, early-type galaxies tend to have older 
ages than late-type and irregular galaxies, and the age differences among
the different morphological types seem to be larger for less massive galaxies.
  In Figure \ref{fig:mwage_morph}, we also show the mass-weighted mean ages
  estimated from both the photometric SED and optical spectrum from the
  LEGA-C survey for galaxies at $z=$ 0.5--1.0 as open symbols.
  We can use total 830 morphologically classified galaxies with 
  the LEGA-C spectra at $z=$ 0.5--1.0, and the numbers of those galaxies
  in each stellar mass and redshift bin are summarised in Table \ref{tab:spec}.
  While median values of the mean ages estimated from both the photometric and
  spectroscopic data are generally consistent with those estimated from only
  the photometric SED, they tend to be older than those estimated from
  only the photometric SED for early-type galaxies, especially, more massive
  ones at lower redshifts. We could underestimate the mean stellar ages of
  those massive and/or lower-$z$ early-type galaxies by $\sim 1$ Gyr,
  and the differences 
  in the mean ages between early-type and late-type/irregular galaxies may be 
  larger at low redshifts.

On the other hand, 
the mean stellar ages of early-type galaxies tends to be 
younger than those of late-type and irregular galaxies at $z\gtrsim1.3$.
The median values for those early-type galaxies become $\lesssim 2$ Gyr
at $z>1.3$, while their dispersions around the median values are relatively
large ($\sim$ 1.0--2.5 Gyr).
Some of those galaxies with young ages may have formed a relatively large
fraction of their stars in the recent past.
While more massive galaxies may have slightly older ages, 
the stellar mass dependence is rather weak in each morphological type.

\section{Discussion} \label{sec:dis}

In this study, we investigated the evolution of the fraction of 
early-type galaxies with $M_{\rm star} > 10^{10.5} M_{\odot}$ at $0.2<z<2.0$, 
by using the $HST$/ACS data and the $JWST$/NIRCam data
over the 0.28 deg$^{2}$ area in the COSMOS field.
We measured the concentration and asymmetry indices $C$ and $A$ on the data, 
and selected early-type galaxies with the criteria of $C > C_{n=2.5}$
and $A_{\rm cor} < 0.2$.

We found that the early-type fraction decreases with increasing redshift
from $\sim$ 40--70\% at $z \sim 0.3$ to $\sim$ 15--25\% at $z\sim 1.8$
depending on stellar mass.
The similar evolution of the early-type fraction has been reported by
previous studies
(\citealp{kaj01}; \citealp{kaj05}; \citealp{bun05}; \citealp{wuy11}; 
\citealp{bel12};
\citealp{bui13}; \citealp{tal14}; \citealp{bru14};
\citealp{dav17}; \citealp{lee24}). 
Several previous studies also found that the median $C$ or S{\'e}rsic index of
massive galaxies similarly decreases with increasing redshift
(\citealp{van10}; 
\citealp{pat13}; \citealp{van13};
\citealp{mor14}; \citealp{pap15}; \citealp{gu19}; 
\citealp{ren24}).
The early-type fraction tends to be higher for more massive galaxies, 
and the fraction in massive galaxies with $M_{\rm star} > 10^{11} M_{\odot}$
increases with time from $\sim$ 20--25\% at $z \sim 1.8$ to $\sim$ 70\% at
$z \sim 0.3$.
\cite{bui13} used the similar sample selection and obtained  
 the quantitatively consistent results for those massive galaxies at $z<3$.
 At $z\sim 1.8$, the early-type fraction is similarly low over
 the stellar mass range we investigated. The larger samples over a wider
 stellar mass range is needed to confirm whether the early-type fraction
 depends on $M_{\rm star}$ or not at such high redshifts.

Instead of early-type galaxies, 
the fraction of irregular galaxies with $A_{\rm cor} > 0.2$
becomes relatively high at $z \gtrsim 1$   
(Figure \ref{fig:frac_morph}). 
This is probably because galaxy interactions/mergers
occurred more frequently at higher redshifts (e.g., \citealp{kaj01};
\citealp{con05}; \citealp{lot11}; \citealp{whi21}). 
In order to reveal how the early-type morphologies with highly 
concentrated surface brightness profiles have formed, 
we investigated the redshift evolution and stellar mass dependence of 
$R_{20}$ and $R_{80}$ for sample galaxies, and found that 
the median $R_{80}$ more strongly depends on $M_{\rm star}$, and significantly
increases with time by $\sim $ 0.2--0.25 dex from $z \sim 1.8$ to 
$z \sim 0.3$.
While the strength of the evolution does not strongly depend on
$M_{\rm star}$ (Figure \ref{fig:r2080_z}),
the combination of this evolution and the strong mass dependence
of the median $R_{80}$ 
 suggests that $R_{80}$ tends to grow significantly as
 $M_{\rm star}$ increases with time.
We note that the progenitor bias could contribute to the evolution of $R_{80}$,
 if galaxies whose stellar mass has reached to the mass limit of
our sample at later epoch tend to have larger $R_{80}$ (e.g., \citealp{car13}). 
On the other hand, the median $R_{20}$ shows no significant mass dependence
and redshift evolution over $0.2<z<2.0$.
\cite{van10} and \cite{pat13} performed stacking analyses of the surface
brightness profiles for massive galaxies with
fixed cumulative number densities, 
and found that mass growth of those galaxies at $z \lesssim 2$ mainly occurred  
in their outer regions.
\cite{mor15} reported the similar inside-out stellar mass growth for
massive galaxies in their stacking analysis of the stellar mass radial profile,
while they also found that less massive galaxies self-similarly increase 
their stellar mass irrespective of radius. 
Our results are also consistent with such inside-out stellar mass 
growth of massive galaxies since $z \sim 2$.
We can expect that the significant growth of $R_{80}$ with nearly constant
$R_{20}$ increases the fraction of those galaxies with high $C$ values,
and the strong mass dependence of $R_{80}$ causes 
the mass dependence of the early-type fraction.
Previous theoretical studies suggested that such stellar mass growth 
in the outer regions of galaxies could be driven by minor merger,
where accreted stars could be preferentially located in the outer regions 
(e.g., \citealp{naa09}; \citealp{ose10}; \citealp{hil12}; \citealp{kar19};
\citealp{rem22}).
The inside-out stellar mass growth by in-situ star formation has been also
suggested by previous observational studies of spatially resolved star
formation activities in star-forming galaxies at $z \lesssim $ 2 
(e.g., \citealp{wuy11}; \citealp{wuy13}; \citealp{nel16}; \citealp{abd18};
\citealp{wil20}).

\begin{figure}
  \includegraphics[width=\columnwidth]{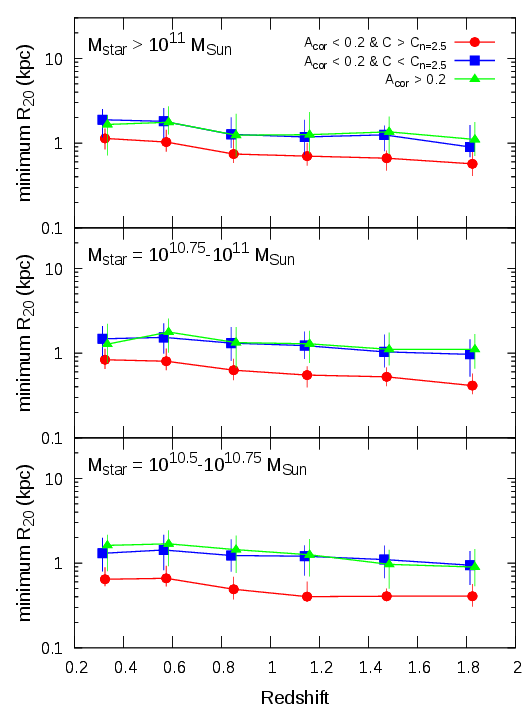}
  \caption{
    Evolution of the minimum $R_{20}$, which is estimated by searching
    a aperture centre that minimises $R_{20}$ within the object, 
    for sample galaxies with the different morphological types
    and stellar masses.
    The symbols are the same as Figure \ref{fig:r2080_morph}.
  \label{fig:minr20}}
\end{figure}

The dispersion around the median values of $R_{20}$ 
is larger than that of $R_{80}$, and the variations among the different
morphological types are larger in $R_{20}$ (Figures \ref{fig:r2080_z} and
\ref{fig:r2080_morph}).
Early-type galaxies have systematically smaller $R_{20}$ than 
late-type and irregular galaxies by 0.25--0.45 and 0.3--0.6 dex, respectively.
Their small $R_{20}$ indicates high surface brightness/stellar mass density
in the central region of the galaxy.
For example, a galaxy with $R_{20} = $ 1 kpc, $M_{\rm star} = 10^{11} M_{\odot}$,
and apparent axial ratio of $b/a = 0.7$ are expected to have 
a central surface stellar mass density of
$\Sigma_{20} = 0.2\times M_{\rm star}/\pi R_{20}^{2} \sim 9 \times 10^{9}$
$M_{\odot}/{\rm kpc}^{2}$, if we assume no radial gradient in the $M/L$ ratio
for the rest-frame $V$ to $R$ band.
The characteristic small $R_{20}$ of early-type galaxies suggests that
growing the central stellar mass density is essential for building 
the early-type morphology.
Several theoretical studies proposed that wet major/minor mergers,
tidal interactions, counter-rotating cold streams, and so on, could cause 
gas infall into the centre of galaxies, which leads to nuclear starburst
and rapid growth of the central stellar mass density (e.g, \citealp{mih96}; 
\citealp{hop06}; \citealp{zol15}; \citealp{man17}; \citealp{lap23}).
Giant clumps formed in a gravitationally unstable, turbulent gas disk
at high redshifts could migrate into the centre, which also triggers
the growth of the central stellar core/bulge (e.g, \citealp{cev10}).
These processes could form the dense stellar core with
remaining an underlying extended stellar component.
\begin{figure}
  \includegraphics[width=1.0\columnwidth]{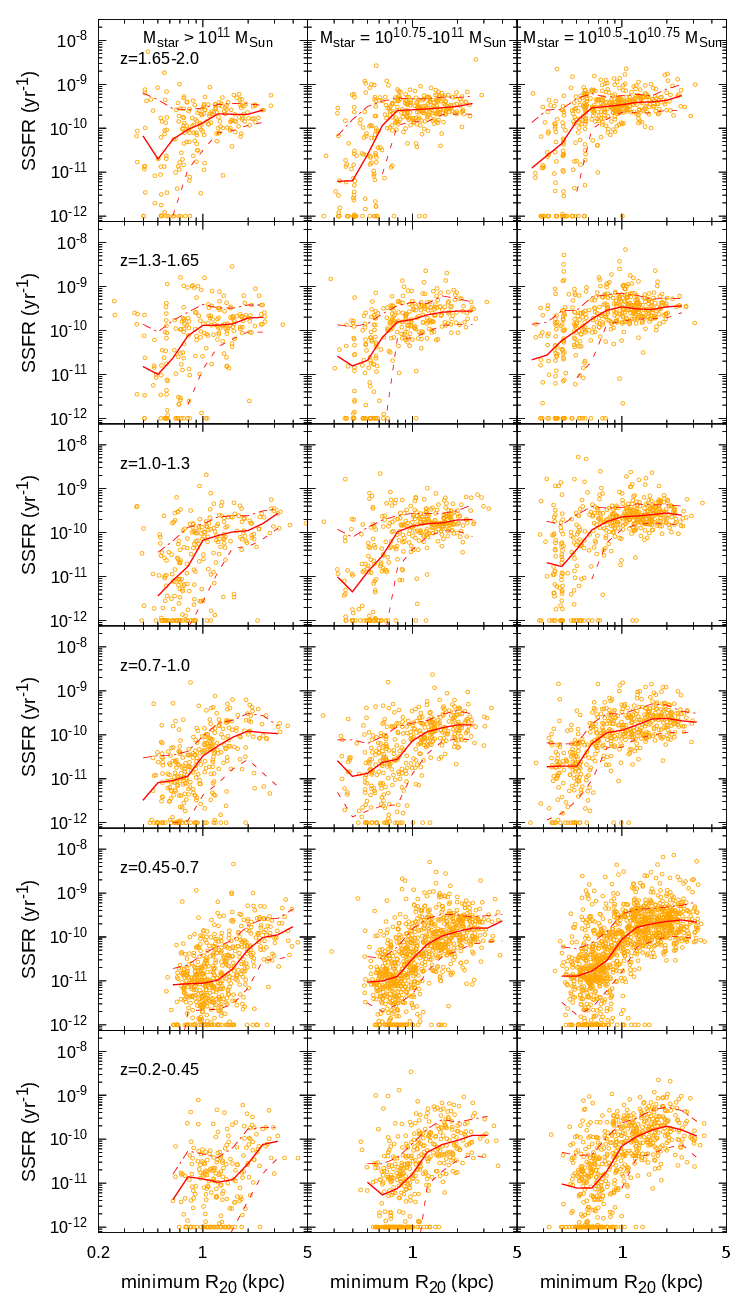}
  \caption{
    SSFR$_{\rm 0-40Myr}$ as a function of minimum $R_{20}$ for sample galaxies
    in the different stellar mass and redshift bins.
    Redshift decreases from the top panels to the bottom panels, while
    stellar mass decreases from the left panels to the right panels.
    The solid line shows median SSFR$_{\rm 0-40Myr}$ in each minimum $R_{20}$
    bin with a width of $\pm 0.1$ dex, while the dashed lines represent
    16 and 84 percentiles of SSFR$_{\rm 0-40Myr}$.
    Note that those galaxies with SSFR$_{\rm 0-40Myr} < 10^{-12}$ yr$^{-1}$ are
    plotted at SSFR$_{\rm 0-40Myr} = 10^{-12}$ yr$^{-1}$.
  \label{fig:ssfr_minR20}}
\end{figure}

\begin{figure}
  \includegraphics[width=1.0\columnwidth]{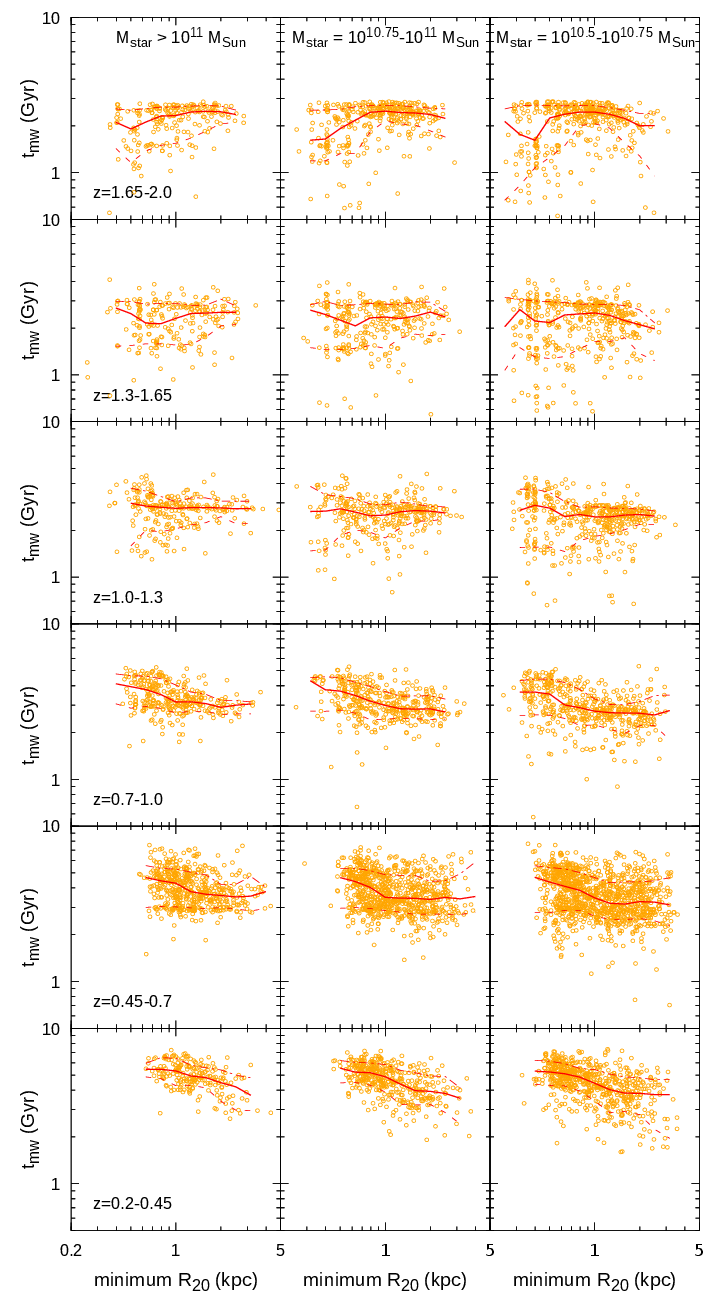}
  \caption{
    The same as Figure \ref{fig:ssfr_minR20} but for mass-weighted mean
    age, $t_{\rm mw}$.
  \label{fig:mwage_minR20}}
\end{figure}

Irregular galaxies with $A_{\rm cor} > 0.2$ in our sample have larger $R_{20}$
than both early-type and late-type galaxies (Figure \ref{fig:r2080_morph}).
Since outer asymmetric features of these galaxies could affect a 
centre position of the aperture used in the $R_{20}$ measurement,
$R_{20}$ can be overestimated for these irregular galaxies if
the centre of the aperture is offset from the dense stellar core.
In order to mitigate such effects of the asymmetric features on $R_{20}$,
we estimated minimum $R_{20}$ by searching the centre position that
minimises $R_{20}$ for sample galaxies.
Figure \ref{fig:minr20} shows minimum $R_{20}$ as a function of redshift
for sample galaxies with the different morphological types and stellar
masses.
While the minimum $R_{20}$ of early-type and late-type galaxies is
similar with $R_{20}$, the minimum $R_{20}$ of irregular
galaxies tends to be slightly smaller than $R_{20}$
in Figure \ref{fig:r2080_morph}.
As a result, irregular galaxies have similar minimum $R_{20}$ with
late-type galaxies, but their minimum $R_{20}$ is significantly larger
than that of early-type galaxies.
Thus most of these irregular galaxies may have not yet formed the dense stellar
core, although we cannot rule out the possibility that such dense core 
is forming with heavy dust extinction.

Regarding on-going star formation activity, we found that early-type galaxies
have $\sim 1$ dex lower SSFRs than late-type and irregular galaxies
irrespective of stellar mass and redshift (Figure \ref{fig:ssfr_morph}), 
which is consistent with previous studies
(e.g., \citealp{wuy11}; \citealp{bel12}; \citealp{lee13}; \citealp{bru14}).
Many previous studies also reported close relationship between 
the surface stellar mass density within 1 kpc from the centre,
$\Sigma_{\rm 1kpc}$ and SSFR of galaxies, namely, "L"-shape distribution 
in the SSFR-$\Sigma_{\rm 1kpc}$ diagram,  
which suggests that the high central stellar mass density seems to be required
for the quenching of star formation in galaxies
(e.g., \citealp{che12}; \citealp{fan13}; \citealp{bar17}; \citealp{mos17};
\citealp{whi17}; \citealp{lee18}).
In Figure \ref{fig:ssfr_minR20}, we show SSFR$_{\rm 0-40Myr}$ as a function of
minimum $R_{20}$ for sample galaxies in the different stellar mass and
redshift bins.
While most of galaxies are star-forming at large minimum $R_{20}$, 
the median values of SSFR$_{\rm 0-40Myr}$ rapidly decreases at minimum
$R_{20} \lesssim 1$ kpc.
The strong minimum $R_{20}$ dependence of SSFR$_{\rm 0-40Myr}$
seems to be consistent with the close relationship between $\Sigma_{\rm 1kpc}$
and SSFR found in the previous studies, because $R_{20}$ is directly related
with the surface stellar mass density within $R_{20}$, $\Sigma_{20}$
as mentioned above.
The transition minimum $R_{20}$, where the median SSFR$_{\rm 0-40Myr}$ rapidly
changes, seems to depend on both stellar mass and redshift, and  
increases with increasing stellar mass and decreasing redshift.
Such mass dependence of the transition $R_{20}$ corresponds to 
a nearly constant transition $\Sigma_{20}$ over the stellar mass range
we investigated,  
while the transition $\Sigma_{20}$ seems to gradually decrease with
decreasing redshift from $\Sigma_{20} \sim 10^{10} M_{\odot}/{\rm kpc}^{2}$
at $z\sim1.8$ to $\sim 3\times 10^{9}  M_{\odot}/{\rm kpc}^{2}$ at $z\sim 0.3$.
While the nuclear starburst mentioned above could rapidly form the dense
stellar core,  
the star formation in the central region might be quenched at a certain
stellar mass density (e.g., \citealp{lap23}).

We found that early-type galaxies have older mass-weighted ages than
late-type and irregular galaxies at $z \lesssim 1$, but early-type galaxies
tend to be younger than late-type and irregular ones at $z \gtrsim 1.3$
(Figure \ref{fig:mwage_morph}).
\citet{tac22} and \citet{ner25} estimated mass-weighted stellar ages of
massive galaxies at $z \sim 0.8$ with high-S/N continuum spectra and
multi-band photometry, and found that massive quiescent galaxies have
ages of $t_{\rm mw} \sim$ 2--6 Gyr with a median value of $\sim$ 4.5 Gyr.
In Figure \ref{fig:mwage_morph}, early-type galaxies at $0.7<z<1.0$
show similar or slightly younger median values of $t_{\rm mw}$.
Their mean stellar ages estimated from both the photometric SEDs and
  LEGA-C spectra (open symbols in the figure) seem to be more consistent with
  these previous studies. The mean ages estimated from only the photometric
  SEDs could be slightly underestimated.
\citet{tac22} and \citet{ner25} also reported that the mass-weighted ages 
of quiescent galaxies are $\sim 1$ Gyr older than those of
star-forming galaxies on average with a large overlap between the two  
populations. In this study, early-type galaxies at $0.7<z<1.0$ are
similarly older than late-type and irregular galaxies,
 although our results from the photometric SED fitting could slightly
  underestimate the age differences among the different morphological types
  as mentioned above.
\citet{slo24} used JWST/NIRSpec deep spectroscopic data to estimate
mass-weighted ages of 20 massive quiescent galaxies at $1<z<3$.
They found that those quiescent galaxies have $t_{\rm mw} \sim $ 0.8--3.0 Gyr,
which is also consistent with early-type galaxies at $1<z<2$ in Figure
\ref{fig:mwage_morph}.

Figure \ref{fig:mwage_minR20} shows mass-weighted mean age as a function
of minimum $R_{20}$ for sample galaxies in the different stellar mass and
redshift bins.
As seen in Figure \ref{fig:mwage_morph}, those galaxies with 
minimum $R_{20} \lesssim 1$ kpc (mainly early-type galaxies) tend to
have younger stellar age at $z \gtrsim 1.3$,
while the dispersion around the median is relatively large.
The young stellar ages of galaxies with minimum $R_{20} \lesssim 1$ kpc
may indicate that the formation of the early-type 
morphology at high redshifts was associated with strong starburst in
the recent past, and such process has started to occur frequently   
since slightly higher redshifts, for example, $z \sim $ 3.
The recent starburst could make the mass-weighted ages of those galaxies
younger, while massive late-type and irregular galaxies may have continued
star formation with a constant or declining SFR for a relatively
long time.

At $z\lesssim 1$, those galaxies with a small minimum $R_{20}$ 
become older than those with larger $R_{20}$ as time elapsed.
In contrast to those galaxies at $z \gtrsim 1.3$, 
there are few early-type galaxies with young ages of 
$t_{\rm mw} < 2$ Gyr at $z < 1$ in Figure \ref{fig:mwage_morph}.
It suggests that 
the formation of the early-type morphology at low redshifts is driven by 
minor mergers or gas-poor major mergers, where associated starburst (if exist)
is not so strong relative to the existing stellar mass (e.g, \citealp{bel06};
\citealp{son14}; \citealp{hai15}; \citealp{can23}).
The weaker contribution of starburst in the morphological transition at
lower redshifts could be explained by the observed evolution of
gas mass fraction of massive galaxies (e.g., \citealp{tac20}).
Those early-type galaxies at $z<1$ with the similar ages as late-type
and irregular ones at the same redshifts may form their morphology
in the relatively recent past by such processes, 
while those galaxies with older ages probably have formed their 
dense stellar core and quenched star formation at higher redshifts.
Such scenario is consistent with the previous studies with high-S/N spectra,
which reported that massive quiescent galaxies at higher redshifts 
tend to have earlier formation epoch and shorter star formation timescale
(e.g., \citealp{gal14}; \citealp{tac22}; \citealp{kau24}; \citealp{bev24}).

\begin{figure}
  \includegraphics[width=1.02\columnwidth]{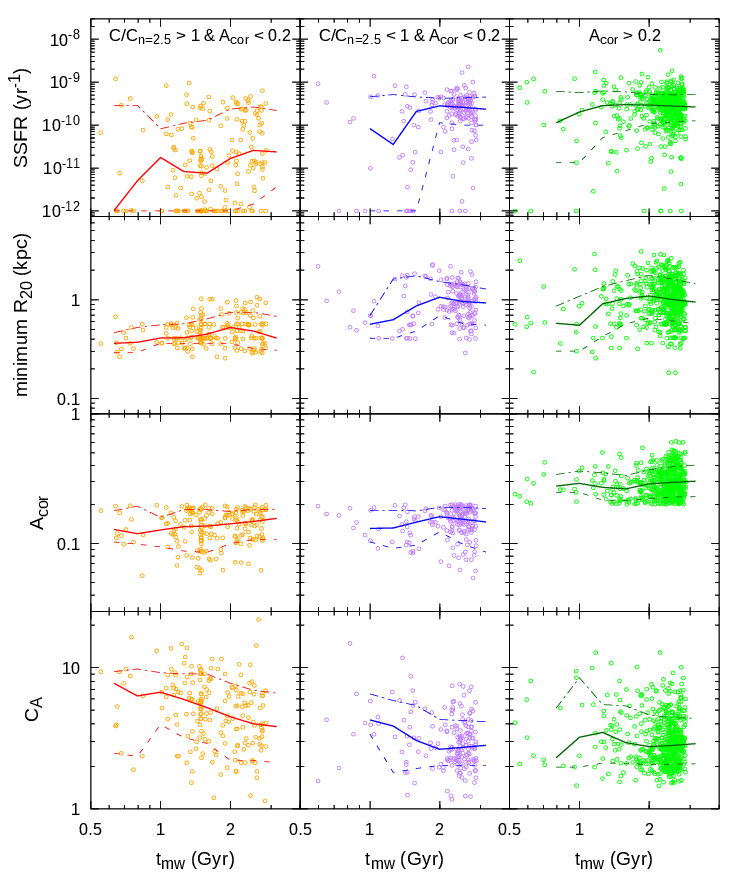}
  \caption{
    SSFR$_{\rm 0-40Myr}$, minimum $R_{20}$, $A$, and $C_{A}$ as a function of
    $t_{\rm mw}$ for sample galaxies with the different morphological
    types at $1.65<z<2.0$.
    The solid line shows a median values in the age bins with 
    a width of $\pm 0.1$ dex, while the dashed lines represent 16 and 84
    percentiles.
    Note that those galaxies with SSFR$_{\rm 0-40Myr} < 10^{-12}$ yr$^{-1}$ are
    plotted at SSFR$_{\rm 0-40Myr} = 10^{-12}$ yr$^{-1}$ in the top panels.
  \label{fig:mor_z16520}}
\end{figure}
\begin{figure*}
  \includegraphics[width=2.0\columnwidth]{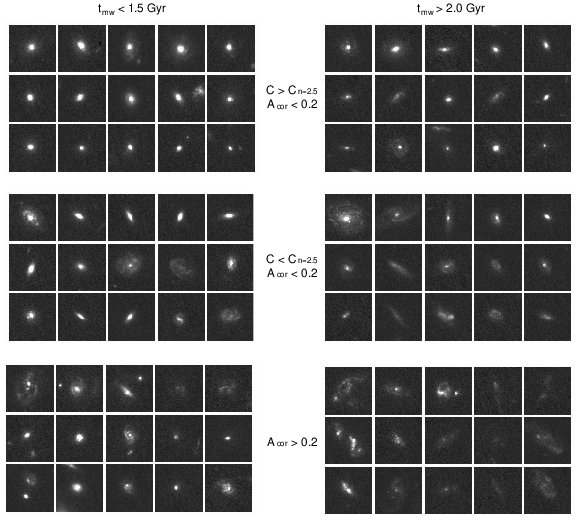}
  \caption{
    F150W-band 3'' $\times$ 3'' images for sample galaxies with
    the different morphological types and mean stellar ages at
    $1.65<z<2.0$.
    Those galaxies with $t_{\rm mw} < 1.5$ Gyr are shown in the left,  
    and those with $t_{\rm mw} > 2$ Gyr are shown in the right.
    Randomly selected 15 galaxies are shown   
    in each morphological type and mean age bin, where 
    $M_{\rm star}$ decreases from top left to bottom right.
  \label{fig:mon}}
\end{figure*}

In the top panels of Figure \ref{fig:mor_z16520}, we show
SSFR$_{\rm 0-40Myr}$ vs. mass-weighted mean age for sample galaxies
with the different morphological types at $1.65<z<2.0$.
The median SSFR$_{\rm 0-40Myr}$ of early-type galaxies depends on
mean stellar age, while the dispersion around the median is relatively large.
The median SSFR$_{\rm 0-40Myr}$ is $\sim 10^{-11}$ yr$^{-1}$ at
$t_{\rm mw} < 1.5$ Gyr, and increases with increasing mean stellar age
at $t_{\rm mw} > 1.5$ Gyr.
Thus those early-type galaxies with young ages tend to be quenched,
although $19 \pm 5$\% of those galaxies show
SSFR$_{\rm 0-40Myr} > 10^{-10}$ yr$^{-1}$.
On the other hand, early-type galaxies with $t_{\rm mw} > 2$ Gyr show
the median SSFR$_{\rm 0-40Myr}$ of $\sim 10^{-10.6}$ yr$^{-1}$, and
$37 \pm 7$\% of those galaxies have SSFR$_{\rm 0-40Myr} > 10^{-10}$ yr$^{-1}$.
Non-negligible star formation occurs in a significant fraction of 
these galaxies with $t_{\rm mw} > 2$ Gyr, but their SSFRs are systematically
lower than those of late-type and irregular galaxies.
Thus those galaxies may have quenched star formation when the dense stellar
core was rapidly formed, and then be rejuvenated about 1--2 Gyr after
the quenching. 
The second to bottom panels of Figure \ref{fig:mor_z16520} show
the morphological indices, namely, minimum $R_{20}$, $A$, and $C_{A}$
as a function of mean stellar age.
While minimum $R_{20}$ and $A_{\rm cor}$ of early-type galaxies slightly
increases with increasing mean stellar age, $C_{A}$ significantly
decreases with increasing mean age at $t_{\rm mw} > 1$ Gyr.
Those early-type galaxies with $t_{\rm mw} < 1.5$ Gyr at $1.65<z<2.0$
have the median values of minimum $R_{20} \sim 0.6$ kpc,
$A_{\rm cor} \sim 0.12$, and $C_{A} \sim $ 6--7.
These properties including low SSFR$_{\rm 0-40Myr}$ and high $C$ are 
similar with compact post-starburst galaxies at $z\sim 0.8$
seen in \cite{him23}. Their high $C_{A}$ values may reflect the
rapid formation of the dense stellar core by intense nuclear starburst 
in the recent past.

On the other hand, early-type galaxies with $t_{\rm mw} > 2$ Gyr
have slightly larger $R_{20}$ and higher $A_{\rm cor}$, and lower $C_{A}$ than
those with $t_{\rm mw} < 1.5$ Gyr.
Their slightly higher $A_{\rm cor}$ values are consistent with the relatively
high SSFR$_{\rm 0-40Myr}$ seen in the top panel,  
if their asymmetric features are associated with star formation activities
such as randomly distributed star-forming regions on a disk.
The median $C_{A}$ values of these galaxies are similar with or slightly
higher than those of late-type and irregular galaxies, which suggests that
the asymmetric features are extended to outer regions of the object.
Figure \ref{fig:mon} shows F150W-band images of sample galaxies with 
the different morphological types and mean stellar ages at $1.65<z<2.0$.
Some of those early-type galaxies with $t_{\rm mw} > 2$ Gyr
have morphologies similar with 
early-type spirals, which have a bright central bulge-like component
and extended (spiral-like) asymmetric features.
These galaxies could have once quenched star formation
when the dense stellar core was formed, and then start forming stars again  
on the extended disk, for example, $\sim $ 1--2 Gyr after the quenching as
mentioned above.
\cite{man19} reported such green-valley galaxies with a very old quenched
bulge and a star-forming disk at $z=$ 0.45--1.
Several theoretical studies with numerical simulations also predicted
the similar rejuvenated early-type disk galaxies that have experienced
the quenching in the bulge formation and restart star formation on a disk later
(e.g., \citealp{spa17}; \citealp{lap23}; \citealp{hop23}).

Late-type galaxies with $C<C_{n=2.5}$ and $A_{\rm cor} < 0.2$ at $1.65<z<2.0$
are basically star-forming galaxies with
SSFR$_{\rm 0-40Myr} \gtrsim 10^{-10}$ yr$^{-1}$, but those galaxies with
$t_{\rm mw} < 1.5$ Gyr show lower median SSFRs of $\sim 10^{-10}$ yr$^{-1}$. 
While the fraction of those with young stellar ages of $t_{\rm mw} < 1.5$ Gyr
is small in late-type galaxies, 
some of those galaxies may be quenching star formation after
the starburst like early-type galaxies with young ages.
Those late-type galaxies with $t_{\rm mw} < 1.5$ Gyr have smaller median
values of minimum $R_{20}$ ($\sim $ 0.6--0.7 kpc) 
than those with $t_{\rm mw} > 2$ Gyr ($\sim$ 1 kpc).
These galaxies also have slightly lower median values of
$A_{\rm cor} \sim 0.12$ and 
higher median values of $C_{A} \sim 4$ than those with older ages.
These morphological properties are similar with those of early-type galaxies
at the same redshifts, in particular, those galaxies with older ages.
In Figure \ref{fig:mon}, late-type galaxies with $t_{\rm mw} < 1.5$ Gyr 
tend to have a bright bulge-like component, while most of those galaxies 
with older ages do not show such a significant bulge. 
Some of those galaxies with young stellar ages could form their bulge
component by starburst in the recent past.

Most of irregular galaxies with $A_{\rm cor} > 0.2$ at $1.65<z<2.0$
have SSFR$_{\rm 0-40Myr}$ and morphological properties different from
early-type galaxies at the same redshifts (Figure \ref{fig:mor_z16520}).
Only a small fraction ($\sim$ 10\%) of irregular galaxies have
$t_{\rm mw} < 1.5$ Gyr, and those irregular galaxies with young stellar ages
have smaller minimum $R_{20}$ and slightly lower SSFRs.
While such young irregular galaxies with small minimum $R_{20}$ could be
immediate progenitors of early-type galaxies,
most irregular galaxies with $t_{\rm mw} \gtrsim 2$ Gyr do not seem to have
a clear bulge-like structure.
The formation of the dense stellar core may have been heavily obscured by dust,
or a significant fraction of asymmetric features may disappear until
the dense core have formed.

\section{Summary}

In order to understand how the early-type morphology formed in galaxies, 
we measured the non-parametric morphological indices, namely,
$C$ ($= R_{80}/R_{20}$), $A$, minimum $R_{20}$, and $C_{A}$  
for galaxies with $M_{\rm star} > 10^{10.5} M_{\odot}$ at $0.2<z<2.0$
by using the $JWST$/NIRCam data from COSMOS-Web survey as well as
$HST$/ACS ones in the COSMOS field. 
We calculated $C_{n=2.5}$, which is a $C$ value of 
the S{\'e}rsic profile with $n=2.5$ for given size and surface brightness
of the object, and corrected $A$ for the resolution effects.
We then investigated the fraction of early-type galaxies
with $C > C_{n=2.5}$ and $A_{\rm cor} < 0.2$ as a function of redshift.
Our main results are summarised as follows.
\begin{itemize}

\item The fraction of early-type galaxies 
  decreases with increasing redshift from $\sim$ 40--70\% at $z \sim 0.3$
  to $\sim$ 15--25\% at $z \sim 1.8$ depending on stellar mass.
  The early-type fraction is higher for more massive galaxies
  at least at $z < 1.65$.

\item While the median $R_{20}$ shows no significant mass dependence
  and redshift evolution, that of $R_{80}$ strongly depends on $M_{\rm star}$
  and significantly increases with time 
  by $\sim $ 0.2--0.25 dex from $z \sim 1.8$ to $z \sim 0.3$.
  The strong growth of $R_{80}$ with nearly constant $R_{20}$ is expected
  to drive the evolution of the early-type fraction.

\item On the other hand, the dispersion around the median values of $R_{20}$
  is larger than that of $R_{80}$, and early-type galaxies have
  characteristically smaller $R_{20}$ than late-type and irregular galaxies.
  Their median $R_{20}$ is smaller than those of late-type and irregular
  galaxies by $\sim $ 0.2--0.5 dex even if we use minimum $R_{20}$,
  which suggests higher stellar mass density in the central region of
  those early-type galaxies.

\item Early-type galaxies have systematically lower SSFR$_{\rm 0-40Myr}$
  than late-type and irregular galaxies by $\sim 1$ dex.
  The median SSFR$_{\rm 0-40Myr}$ of sample galaxies strongly depends
  on $R_{20}$ and rapidly changes around $R_{20} \sim 1$ kpc.
  The formation of the dense stellar core could be closely related with
  the quenching of star formation.

\item While early-type galaxies have older mass-weighted mean stellar ages
  than late-type and irregular galaxies at $z\lesssim 1$,
  the median ages of early-type galaxies become $\lesssim$ 2Gyr at
  $z\gtrsim 1.3$,  which is younger than late-type and irregular galaxies.
  Their young ages suggest that the formation of the early-type morphology
  at high redshifts is associated with strong starburst.

\item Early-type galaxies with $t_{\rm mw} < 1.5$ Gyr at $1.65<z<2.0$
  show lower SSFR$_{\rm 0-40Myr}$ and higher $C_{A}$ than those galaxies with
  $t_{\rm mw} > 2$ Gyr. Those early-type galaxies seem to have quenched 
  star formation when the dense stellar core was rapidly formed, 
  and then some of them may be rejuvenated about $\sim $ 1--2 Gyr
  after the quenching.

\end{itemize}

\section*{Acknowledgements}

We thank the anonymous referee very much
for the valuable suggestions and comments.
This research is based in part on data collected at Subaru Telescope,
which is operated by the National Astronomical Observatory of Japan.
We are honoured and grateful for the opportunity of observing the 
Universe from Maunakea, which has the cultural, historical and natural 
significance in Hawaii.
Based on data products from observations made with ESO Telescopes at the
La Silla Paranal Observatory under programme IDs 194A.2005 and 1100.A-0949
(The LEGA-C Public Spectroscopy Survey). The LEGA-C project has received
funding from the European Research Council (ERC) under the European Unions
Horizon 2020 research and innovation programme (grant agreement No. 683184).
Data analysis were in part carried out on the Multi-wavelength Data Analysis
System operated by the Astronomy Data Center (ADC), National Astronomical
Observatory of Japan.

\section*{Data Availability}

The COSMOS2020 catalogue is publicly available at
https://cosmos2020.calet.org/.
The reduced {\it JWST}/NIRCam data from COSMOS-Web survey are publicly available at 
https://ariel.astro.illinois.edu/cosmos\_web/
The COSMOS {\it HST}/ACS $I_{\rm F814W}$-band mosaic data version 2.0
are publicly available via NASA/IPAC Infrared Science Archive at
https://irsa.ipac.caltech.edu/data/COSMOS/images/acs\_mosaic\_2.0/.
The raw data for the ACS mosaic are available via
Mikulski Archive for Space Telescopes at
https://archive.stsci.edu/missions-and-data/hst.
The zCOSMOS spectroscopic redshift catalogue is publicly available via
ESO Science Archive Facility at
https://www.eso.org/qi/catalog/show/65.
The LEGA-C catalogue is also publicly available at 
https://www.eso.org/qi/catalogQuery/index/379,
and the reduced spectra are available via ESO Science Archive Portal at
https://archive.eso.org/scienceportal/home.
The VIMOS Ultra Deep Survey catalogue is publicly available at
https://cdsarc.cds.unistra.fr/viz-bin/cat/J/A+A/600/A110.
The FMOS-COSMOS catalogue is publicly available at
https://cdsarc.cds.unistra.fr/viz-bin/cat/J/ApJS/220/12.
The COSMOS 10K DEIMOS spectroscopic catalogue is publicly available at 
https://cdsarc.cds.unistra.fr/viz-bin/cat/J/ApJ/858/77.
 




\begin{thebibliography}{99}

\bibitem[\protect\citeauthoryear{Abdurro'uf \& Akiyama}{2018}]{abd18} Abdurro'uf, Akiyama M., 2018, MNRAS, 479, 5083
  
\bibitem[\protect\citeauthoryear{Aihara et al.}{2019}]{aih19} Aihara H., AlSayyad Y., Ando M., Armstrong R., Bosch J., Egami E., Furusawa H., et al., 2019, PASJ, 71, 114

\bibitem[\protect\citeauthoryear{Ashby et al.}{2018}]{ash18} Ashby M.~L.~N., Caputi K.~I., Cowley W., Deshmukh S., Dunlop J.~S., Milvang-Jensen B., Fynbo J.~P.~U., et al., 2018, ApJS, 237, 39

\bibitem[\protect\citeauthoryear{Ashby et al.}{2013}]{ash13} Ashby M.~L.~N., Willner S.~P., Fazio G.~G., Huang J.-S., Arendt R., Barmby P., Barro G., et al., 2013, ApJ, 769, 80

\bibitem[\protect\citeauthoryear{Ashby et al.}{2015}]{ash15} Ashby M.~L.~N., Willner S.~P., Fazio G.~G., Dunlop J.~S., Egami E., Faber S.~M., Ferguson H.~C., et al., 2015, ApJS, 218, 33

\bibitem[\protect\citeauthoryear{Barro et al.}{2017}]{bar17} Barro G., Faber S.~M., Koo D.~C., Dekel A., Fang J.~J., Trump J.~R., P{\'e}rez-Gonz{\'a}lez P.~G., et al., 2017, ApJ, 840, 47

\bibitem[\protect\citeauthoryear{Beckwith et al.}{2006}]{bec06} Beckwith S.~V.~W., Stiavelli M., Koekemoer A.~M., Caldwell J.~A.~R., Ferguson H.~C., Hook R., Lucas R.~A., et al., 2006, AJ, 132, 1729

\bibitem[\protect\citeauthoryear{Bell et al.}{2006}]{bel06} Bell E.~F., Naab T., McIntosh D.~H., Somerville R.~S., Caldwell J.~A.~R., Barden M., Wolf C., et al., 2006, ApJ, 640, 241
  
\bibitem[\protect\citeauthoryear{Bell et al.}{2012}]{bel12} Bell E.~F., van der Wel A., Papovich C., Kocevski D., Lotz J., McIntosh D.~H., Kartaltepe J., et al., 2012, ApJ, 753, 167
    
\bibitem[\protect\citeauthoryear{Bertin \& Arnouts}{1996}]{ber96} Bertin E., Arnouts S., 1996, A\&AS, 117, 393

\bibitem[\protect\citeauthoryear{Beverage et al.}{2024}]{bev24} Beverage A.~G., Kriek M., Suess K.~A., Conroy C., Price S.~H., Barro G., Bezanson R., et al., 2024, ApJ, 966, 234
  
\bibitem[\protect\citeauthoryear{Bluck et al.}{2019}]{blu19} Bluck A.~F.~L., Bottrell C., Teimoorinia H., Henriques B.~M.~B., Mendel J.~T., Ellison S.~L., Thanjavur K., et al., 2019, MNRAS, 485, 666

\bibitem[\protect\citeauthoryear{Brinchmann et al.}{1998}]{bri98} Brinchmann J., Abraham R., Schade D., Tresse L., Ellis R.~S., Lilly S., Le F{\`e}vre O., et al., 1998, ApJ, 499, 112. doi:10.1086/305621

\bibitem[\protect\citeauthoryear{Brinchmann et al.}{2004}]{bri04} Brinchmann J., Charlot S., White S.~D.~M., Tremonti C., Kauffmann G., Heckman T., Brinkmann J., 2004, MNRAS, 351, 1151
  
\bibitem[\protect\citeauthoryear{Brinchmann \& Ellis}{2000}]{bri00} Brinchmann J., Ellis R.~S., 2000, ApJL, 536, L77

\bibitem[\protect\citeauthoryear{Bruce et al.}{2014}]{bru14} Bruce V.~A., Dunlop J.~S., McLure R.~J., Cirasuolo M., Buitrago F., Bowler R.~A.~A., Targett T.~A., et al., 2014, MNRAS, 444, 1001
  
\bibitem[\protect\citeauthoryear{Bruzual \& Charlot}{2003}]{bru03} Bruzual G., Charlot S., 2003, MNRAS, 344, 1000

\bibitem[\protect\citeauthoryear{Buitrago et al.}{2013}]{bui13} Buitrago F., Trujillo I., Conselice C.~J., H{\"a}u{\ss}ler B., 2013, MNRAS, 428, 1460

\bibitem[\protect\citeauthoryear{Bundy, Ellis, \& Conselice}{2005}]{bun05} Bundy K., Ellis R.~S., Conselice C.~J., 2005, ApJ, 625, 621
  
\bibitem[\protect\citeauthoryear{Calzetti et al.}{2000}]{cal00} Calzetti D., Armus L., Bohlin R.~C., Kinney A.~L., Koornneef J., Storchi-Bergmann T., 2000, ApJ, 533, 682

\bibitem[\protect\citeauthoryear{Cannarozzo et al.}{2023}]{can23} Cannarozzo C., Leauthaud A., Oyarz{\'u}n G.~A., Nipoti C., Diemer B., Huang S., Rodriguez-Gomez V., et al., 2023, MNRAS, 520, 5651

\bibitem[\protect\citeauthoryear{Carollo et al.}{2013}]{car13} Carollo C.~M., Bschorr T.~J., Renzini A., Lilly S.~J., Capak P., Cibinel A., Ilbert O., et al., 2013, ApJ, 773, 112
  
\bibitem[\protect\citeauthoryear{Casey et al.}{2023}]{cas23} Casey C.~M., Kartaltepe J.~S., Drakos N.~E., Franco M., Harish S., Paquereau L., Ilbert O., et al., 2023, ApJ, 954, 31

\bibitem[\protect\citeauthoryear{Ceverino, Dekel, \& Bournaud}{2010}]{cev10} Ceverino D., Dekel A., Bournaud F., 2010, MNRAS, 404, 2151

\bibitem[\protect\citeauthoryear{Chauke et al.}{2018}]{cha18} Chauke P., van der Wel A., Pacifici C., Bezanson R., Wu P.-F., Gallazzi A., Noeske K., et al., 2018, ApJ, 861, 13
  
\bibitem[\protect\citeauthoryear{Chabrier}{2003}]{cha03} Chabrier G., 2003, PASP, 115, 763

\bibitem[\protect\citeauthoryear{Cheung et al.}{2012}]{che12} Cheung E., Faber S.~M., Koo D.~C., Dutton A.~A., Simard L., McGrath E.~J., Huang J.-S., et al., 2012, ApJ, 760, 131

\bibitem[\protect\citeauthoryear{Conroy}{2013}]{con13} Conroy C., 2013, ARA\&A, 51, 393. doi:10.1146/annurev-astro-082812-141017
  
\bibitem[\protect\citeauthoryear{Conselice}{2014}]{con14} Conselice C.~J., 2014, ARA\&A, 52, 291
  
\bibitem[\protect\citeauthoryear{Conselice, Bershady, \& Jangren}{2000}]{con00} Conselice C.~J., Bershady M.~A., Jangren A., 2000, ApJ, 529, 886

\bibitem[\protect\citeauthoryear{Conselice, Blackburne, \& Papovich}{2005}]{con05} Conselice C.~J., Blackburne J.~A., Papovich C., 2005, ApJ, 620, 564

\bibitem[\protect\citeauthoryear{Davari et al.}{2017}]{dav17} Davari R.~H., Ho L.~C., Mobasher B., Canalizo G., 2017, ApJ, 836, 75

\bibitem[\protect\citeauthoryear{Deg et al.}{2023}]{deg23} Deg N., Perron-Cormier M., Spekkens K., Glowacki M., Blyth S.-L., Hank N., 2023, MNRAS, 523, 4340

\bibitem[\protect\citeauthoryear{de Vaucouleurs}{1948}]{dev48} de Vaucouleurs G., 1948, AnAp, 11, 247
  
\bibitem[\protect\citeauthoryear{Driver et al.}{2007}]{dri07} Driver S.~P., Allen P.~D., Liske J., Graham A.~W., 2007, ApJL, 657, L85

\bibitem[\protect\citeauthoryear{Fang et al.}{2013}]{fan13} Fang J.~J., Faber S.~M., Koo D.~C., Dekel A., 2013, ApJ, 776, 63

\bibitem[\protect\citeauthoryear{Ferreira et al.}{2023}]{fer23} Ferreira L., Conselice C.~J., Sazonova E., Ferrari F., Caruana J., Tohill C.-B., Lucatelli G., et al., 2023, ApJ, 955, 94

\bibitem[\protect\citeauthoryear{Franx et al.}{2008}]{fra08} Franx M., van Dokkum P.~G., F{\"o}rster Schreiber N.~M., Wuyts S., Labb{\'e} I., Toft S., 2008, ApJ, 688, 770

\bibitem[\protect\citeauthoryear{Gallazzi et al.}{2014}]{gal14} Gallazzi A., Bell E.~F., Zibetti S., Brinchmann J., Kelson D.~D., 2014, ApJ, 788, 72
  
\bibitem[\protect\citeauthoryear{Giavalisco et al.}{2004}]{gia04} Giavalisco M., Ferguson H.~C., Koekemoer A.~M., Dickinson M., Alexander D.~M., Bauer F.~E., Bergeron J., et al., 2004, ApJL, 600, L93

\bibitem[\protect\citeauthoryear{Gu et al.}{2019}]{gu19} Gu Y., Fang G., Yuan Q., Lu S., Li F., Cai Z.-Y., Kong X., et al., 2019, ApJ, 884, 172

\bibitem[\protect\citeauthoryear{Haines et al.}{2015}]{hai15} Haines T., McIntosh D.~H., S{\'a}nchez S.~F., Tremonti C., Rudnick G., 2015, MNRAS, 451, 433
  
\bibitem[\protect\citeauthoryear{Hasinger et al.}{2018}]{has18} Hasinger G., Capak P., Salvato M., Barger A.~J., Cowie L.~L., Faisst A., Hemmati S., et al., 2018, ApJ, 858, 77

\bibitem[\protect\citeauthoryear{Hilz et al.}{2012}]{hil12} Hilz M., Naab T., Ostriker J.~P., Thomas J., Burkert A., Jesseit R., 2012, MNRAS, 425, 3119
  
\bibitem[\protect\citeauthoryear{Himoto \& Kajisawa}{2023}]{him23} Himoto K.~G., Kajisawa M., 2023, MNRAS, 519, 4110
  
\bibitem[\protect\citeauthoryear{Hopkins et al.}{2008}]{hop08} Hopkins P.~F., Cox T.~J., Kere{\v{s}} D., Hernquist L., 2008, ApJS, 175, 390
  
\bibitem[\protect\citeauthoryear{Hopkins et al.}{2023}]{hop23} Hopkins P.~F., Gurvich A.~B., Shen X., Hafen Z., Grudi{\'c} M.~Y., Kurinchi-Vendhan S., Hayward C.~C., et al., 2023, MNRAS, 525, 2241
  
\bibitem[\protect\citeauthoryear{Hopkins et al.}{2006}]{hop06} Hopkins P.~F., Hernquist L., Cox T.~J., Di Matteo T., Robertson B., Springel V., 2006, ApJS, 163, 1
  
\bibitem[\protect\citeauthoryear{Ilbert et al.}{2010}]{ilb10} Ilbert O., Salvato M., Le Floc'h E., Aussel H., Capak P., McCracken H.~J., Mobasher B., et al., 2010, ApJ, 709, 644
  
\bibitem[\protect\citeauthoryear{Kajisawa \& Yamada}{2001}]{kaj01} Kajisawa M., Yamada T., 2001, PASJ, 53, 833

\bibitem[\protect\citeauthoryear{Kajisawa \& Yamada}{2005}]{kaj05} Kajisawa M., Yamada T., 2005, ApJ, 618, 91

\bibitem[\protect\citeauthoryear{Karademir et al.}{2019}]{kar19} Karademir G.~S., Remus R.-S., Burkert A., Dolag K., Hoffmann T.~L., Moster B.~P., Steinwandel U.~P., et al., 2019, MNRAS, 487, 318

\bibitem[\protect\citeauthoryear{Kauffmann et al.}{2003}]{kau03} Kauffmann G., Heckman T.~M., White S.~D.~M., Charlot S., Tremonti C., Peng E.~W., Seibert M., et al., 2003, MNRAS, 341, 54

\bibitem[\protect\citeauthoryear{Kaushal et al.}{2024}]{kau24} Kaushal Y., Nersesian A., Bezanson R., van der Wel A., Leja J., Carnall A., Gallazzi A., et al., 2024, ApJ, 961, 118
  
\bibitem[\protect\citeauthoryear{Kelvin et al.}{2014}]{kel14} Kelvin L.~S., Driver S.~P., Robotham A.~S.~G., Taylor E.~N., Graham A.~W., Alpaslan M., Baldry I., et al., 2014, MNRAS, 444, 1647
  
\bibitem[\protect\citeauthoryear{Koekemoer et al.}{2007}]{koe07} Koekemoer A.~M., Aussel H., Calzetti D., Capak P., Giavalisco M., Kneib J.-P., Leauthaud A., et al., 2007, ApJS, 172, 196

\bibitem[\protect\citeauthoryear{Kuhn et al.}{2024}]{kuh24} Kuhn V., Guo Y., Martin A., Bayless J., Gates E., Puleo A., 2024, ApJL, 968, L15

\bibitem[\protect\citeauthoryear{Lapiner et al.}{2023}]{lap23} Lapiner S., Dekel A., Freundlich J., Ginzburg O., Jiang F., Kretschmer M., Tacchella S., et al., 2023, MNRAS, 522, 4515
  
\bibitem[\protect\citeauthoryear{Lawson \& Hanson}{1974}]{law74} Lawson C.~L., Hanson R.~J., 1974, slsp.book

\bibitem[\protect\citeauthoryear{Le Conte et al.}{2024}]{lec24} Le Conte Z.~A., Gadotti D.~A., Ferreira L., Conselice C.~J., de S{\'a}-Freitas C., Kim T., Neumann J., et al., 2024, MNRAS, 530, 1984
  
\bibitem[\protect\citeauthoryear{Lee et al.}{2018}]{lee18} Lee B., Giavalisco M., Whitaker K., Williams C.~C., Ferguson H.~C., Acquaviva V., Koekemoer A.~M., et al., 2018, ApJ, 853, 131

\bibitem[\protect\citeauthoryear{Lee et al.}{2013}]{lee13} Lee B., Giavalisco M., Williams C.~C., Guo Y., Lotz J., Van der Wel A., Ferguson H.~C., et al., 2013, ApJ, 774, 47
  
\bibitem[\protect\citeauthoryear{Lee et al.}{2024}]{lee24} Lee J.~H., Park C., Hwang H.~S., Kwon M., 2024, ApJ, 966, 113

\bibitem[\protect\citeauthoryear{Leja et al.}{2017}]{lej17} Leja J., Johnson B.~D., Conroy C., van Dokkum P.~G., Byler N., 2017, ApJ, 837, 170

\bibitem[\protect\citeauthoryear{Lilly et al.}{2009}]{lil09} Lilly S.~J., Le Brun V., Maier C., Mainieri V., Mignoli M., Scodeggio M., Zamorani G., et al., 2009, ApJS, 184, 218

\bibitem[\protect\citeauthoryear{Lilly et al.}{2007}]{lil07} Lilly S.~J., Le F{\`e}vre O., Renzini A., Zamorani G., Scodeggio M., Contini T., Carollo C.~M., et al., 2007, ApJS, 172, 70

\bibitem[\protect\citeauthoryear{Lotz et al.}{2011}]{lot11} Lotz J.~M., Jonsson P., Cox T.~J., Croton D., Primack J.~R., Somerville R.~S., Stewart K., 2011, ApJ, 742, 103

\bibitem[\protect\citeauthoryear{Lotz, Primack, \& Madau}{2004}]{lot04} Lotz J.~M., Primack J., Madau P., 2004, AJ, 128, 163
  
\bibitem[\protect\citeauthoryear{Luo et al.}{2025}]{luo25} Luo Y., Sha A., Ren J., Zhang X., Meng X., Li N., Liu F.~S., 2025, PASP, 137, 064101
  
\bibitem[\protect\citeauthoryear{Mager et al.}{2018}]{mag18} Mager V.~A., Conselice C.~J., Seibert M., Gusbar C., Katona A.~P., Villari J.~M., Madore B.~F., et al., 2018, ApJ, 864, 123
  
\bibitem[\protect\citeauthoryear{Magris et al.}{2015}]{mag15} Magris C.~G., Mateu P.~J., Mateu C., Bruzual A.~G., Cabrera-Ziri I., Mej{\'\i}a-Narv{\'a}ez A., 2015, PASP, 127, 16
  
\bibitem[\protect\citeauthoryear{Mancini et al.}{2019}]{man19} Mancini C., Daddi E., Juneau S., Renzini A., Rodighiero G., Cappellari M., Rodr{\'\i}guez-Mu{\~n}oz L., et al., 2019, MNRAS, 489, 1265

\bibitem[\protect\citeauthoryear{Mancini et al.}{2010}]{man10} Mancini C., Daddi E., Renzini A., Salmi F., McCracken H.~J., Cimatti A., Onodera M., et al., 2010, MNRAS, 401, 933
  
\bibitem[\protect\citeauthoryear{Martig et al.}{2009}]{mar09} Martig M., Bournaud F., Teyssier R., Dekel A., 2009, ApJ, 707, 250
  
\bibitem[\protect\citeauthoryear{Mawatari et al.}{2020}]{maw20} Mawatari K., Inoue A.~K., Yamanaka S., Hashimoto T., Tamura Y., 2020, IAUS, 341, 285

\bibitem[\protect\citeauthoryear{Mawatari et al.}{2016}]{maw16} Mawatari K., Yamada T., Fazio G.~G., Huang J.-S., Ashby M.~L.~N., 2016, PASJ, 68, 46
  
\bibitem[\protect\citeauthoryear{McCracken et al.}{2012}]{mcc12} McCracken H.~J., Milvang-Jensen B., Dunlop J., Franx M., Fynbo J.~P.~U., Le F{\`e}vre O., Holt J., et al., 2012, A\&A, 544, A156

  \bibitem[\protect\citeauthoryear{Mandelker et al.}{2017}]{man17} Mandelker N., Dekel A., Ceverino D., DeGraf C., Guo Y., Primack J., 2017, MNRAS, 464, 635

\bibitem[\protect\citeauthoryear{Mihos \& Hernquist}{1996}]{mih96} Mihos J.~C., Hernquist L., 1996, ApJ, 464, 641

\bibitem[\protect\citeauthoryear{Morishita, Ichikawa, \& Kajisawa}{2014}]{mor14} Morishita T., Ichikawa T., Kajisawa M., 2014, ApJ, 785, 18

\bibitem[\protect\citeauthoryear{Morishita et al.}{2015}]{mor15} Morishita T., Ichikawa T., Noguchi M., Akiyama M., Patel S.~G., Kajisawa M., Obata T., 2015, ApJ, 805, 34
  
\bibitem[\protect\citeauthoryear{Mosleh et al.}{2017}]{mos17} Mosleh M., Tacchella S., Renzini A., Carollo C.~M., Molaeinezhad A., Onodera M., Khosroshahi H.~G., et al., 2017, ApJ, 837, 2

\bibitem[\protect\citeauthoryear{Naab, Johansson, \& Ostriker}{2009}]{naa09} Naab T., Johansson P.~H., Ostriker J.~P., 2009, ApJL, 699, L178
  
\bibitem[\protect\citeauthoryear{Nelson et al.}{2016}]{nel16} Nelson E.~J., van Dokkum P.~G., F{\"o}rster Schreiber N.~M., Franx M., Brammer G.~B., Momcheva I.~G., Wuyts S., et al., 2016, ApJ, 828, 27

\bibitem[\protect\citeauthoryear{Nersesian et al.}{2024}]{ner24} Nersesian A., van der Wel A., Gallazzi A., Leja J., Bezanson R., Bell E.~F., D'Eugenio F., et al., 2024, A\&A, 681, A94
  
\bibitem[\protect\citeauthoryear{Nersesian et al.}{2025}]{ner25} Nersesian A., van der Wel A., Gallazzi A.~R., Kaushal Y., Bezanson R., Zibetti S., Bell E.~F., et al., 2025, A\&A, 695, A86
  
\bibitem[\protect\citeauthoryear{Omand, Balogh, \& Poggianti}{2014}]{omo14} Omand C.~M.~B., Balogh M.~L., Poggianti B.~M., 2014, MNRAS, 440, 843
  
\bibitem[\protect\citeauthoryear{Ormerod et al.}{2024}]{orm24} Ormerod K., Conselice C.~J., Adams N.~J., Harvey T., Austin D., Trussler J., Ferreira L., et al., 2024, MNRAS, 527, 6110
  
\bibitem[\protect\citeauthoryear{Oser et al.}{2010}]{ose10} Oser L., Ostriker J.~P., Naab T., Johansson P.~H., Burkert A., 2010, ApJ, 725, 2312
  
\bibitem[\protect\citeauthoryear{Papovich et al.}{2015}]{pap15} Papovich C., Labb{\'e} I., Quadri R., Tilvi V., Behroozi P., Bell E.~F., Glazebrook K., et al., 2015, ApJ, 803, 26
  
\bibitem[\protect\citeauthoryear{Patel et al.}{2013}]{pat13} Patel S.~G., van Dokkum P.~G., Franx M., Quadri R.~F., Muzzin A., Marchesini D., Williams R.~J., et al., 2013, ApJ, 766, 15

\bibitem[\protect\citeauthoryear{Remus \& Forbes}{2022}]{rem22} Remus R.-S., Forbes D.~A., 2022, ApJ, 935, 37
  
\bibitem[\protect\citeauthoryear{Ren et al.}{2024}]{ren24} Ren J., Liu F.~S., Li N., Cui Q., Zhao P., Li Y., Song Q., et al., 2024, ApJ, 969, 4
  
\bibitem[\protect\citeauthoryear{Roberts \& Haynes}{1994}]{rob94} Roberts M.~S., Haynes M.~P., 1994, ARA\&A, 32, 115

\bibitem[\protect\citeauthoryear{Salvador et al.}{2024}]{sal24} Salvador D., Cerulo P., Valenzuela K., Demarco R., Oyarzo F., Gatica C., 2024, A\&A, 684, A166
  
\bibitem[\protect\citeauthoryear{Salim, Boquien, \& Lee}{2018}]{sal18} Salim S., Boquien M., Lee J.~C., 2018, ApJ, 859, 11

\bibitem[\protect\citeauthoryear{Salim \& Narayanan}{2020}]{sal20} Salim S., Narayanan D., 2020, ARA\&A, 58, 529
  
\bibitem[\protect\citeauthoryear{Sawicki et al.}{2019}]{saw19} Sawicki M., Arnouts S., Huang J., Coupon J., Golob A., Gwyn S., Foucaud S., et al., 2019, MNRAS, 489, 5202

\bibitem[\protect\citeauthoryear{Sazonova et al.}{2024}]{saz24} Sazonova E., Morgan C., Balogh M., Alatalo K., Benavides J.~A., Bluck A., Brough S., et al., 2024, OJAp, 7, 77
  
\bibitem[\protect\citeauthoryear{Scoville et al.}{2007a}]{sco07a} Scoville N., Aussel H., Brusa M., Capak P., Carollo C.~M., Elvis M., Giavalisco M., et al., 2007, ApJS, 172, 1

\bibitem[\protect\citeauthoryear{Scoville et al.}{2007b}]{sco07b} Scoville N., Abraham R.~G., Aussel H., Barnes J.~E., Benson A., Blain A.~W., Calzetti D., et al., 2007, ApJS, 172, 38

\bibitem[\protect\citeauthoryear{S{\'e}rsic}{1968}]{ser68} S{\'e}rsic J.~L., 1968, adga.book

\bibitem[\protect\citeauthoryear{Silverman et al.}{2015}]{sil15} Silverman J.~D., Kashino D., Sanders D., Kartaltepe J.~S., Arimoto N., Renzini A., Rodighiero G., et al., 2015, ApJS, 220, 12

\bibitem[\protect\citeauthoryear{Slob et al.}{2024}]{slo24} Slob M., Kriek M., Beverage A.~G., Suess K.~A., Barro G., Bezanson R., Brammer G., et al., 2024, ApJ, 973, 131

\bibitem[\protect\citeauthoryear{Sonnenfeld, Nipoti, \& Treu}{2014}]{son14} Sonnenfeld A., Nipoti C., Treu T., 2014, ApJ, 786, 89
  
\bibitem[\protect\citeauthoryear{Sparre et al.}{2017}]{spa17} Sparre M., Hayward C.~C., Feldmann R., Faucher-Gigu{\`e}re C.-A., Muratov A.~L., Kere{\v{s}} D., Hopkins P.~F., 2017, MNRAS, 466, 88
  
\bibitem[\protect\citeauthoryear{Steinhardt et al.}{2014}]{ste14} Steinhardt C.~L., Speagle J.~S., Capak P., Silverman J.~D., Carollo M., Dunlop J., Hashimoto Y., et al., 2014, ApJL, 791, L25
  
\bibitem[\protect\citeauthoryear{Tacchella et al.}{2015}]{tac15} Tacchella S., Carollo C.~M., Renzini A., F{\"o}rster Schreiber N.~M., Lang P., Wuyts S., Cresci G., et al., 2015, Sci, 348, 314

\bibitem[\protect\citeauthoryear{Tacchella et al.}{2022}]{tac22} Tacchella S., Conroy C., Faber S.~M., Johnson B.~D., Leja J., Barro G., Cunningham E.~C., et al., 2022, ApJ, 926, 134

\bibitem[\protect\citeauthoryear{Tacconi, Genzel, \& Sternberg}{2020}]{tac20} Tacconi L.~J., Genzel R., Sternberg A., 2020, ARA\&A, 58, 157

\bibitem[\protect\citeauthoryear{Talia et al.}{2014}]{tal14} Talia M., Cimatti A., Mignoli M., Pozzetti L., Renzini A., Kurk J., Halliday C., 2014, A\&A, 562, A113
  
\bibitem[\protect\citeauthoryear{Tan et al.}{2024}]{tan24} Tan V.~Y.~Y., Muzzin A., Marchesini D., Sok V., Sarrouh G.~T.~E., Marsan Z.~C., 2024, ApJ, 964, 177
  
\bibitem[\protect\citeauthoryear{Taniguchi et al.}{2007}]{tan07} Taniguchi Y., Scoville N., Murayama T., Sanders D.~B., Mobasher B., Aussel H., Capak P., et al., 2007, ApJS, 172, 9

\bibitem[\protect\citeauthoryear{Taniguchi et al.}{2015}]{tan15} Taniguchi Y., Kajisawa M., Kobayashi M.~A.~R., Shioya Y., Nagao T., Capak P.~L., Aussel H., et al., 2015, PASJ, 67, 104

\bibitem[\protect\citeauthoryear{Tasca et al.}{2017}]{tas17} Tasca L.~A.~M., Le F{\`e}vre O., Ribeiro B., Thomas R., Moreau C., Cassata P., Garilli B., et al., 2017, A\&A, 600, A110

\bibitem[\protect\citeauthoryear{Thorp et al.}{2021}]{tho21} Thorp M.~D., Bluck A.~F.~L., Ellison S.~L., Maiolino R., Conselice C.~J., Hani M.~H., Bottrell C., 2021, MNRAS, 507, 886
  
\bibitem[\protect\citeauthoryear{Tojeiro et al.}{2007}]{toj07} Tojeiro R., Heavens A.~F., Jimenez R., Panter B., 2007, MNRAS, 381, 1252

\bibitem[\protect\citeauthoryear{van den Bergh et al.}{2000}]{van00} van den Bergh S., Cohen J.~G., Hogg D.~W., Blandford R., 2000, AJ, 120, 2190
  
\bibitem[\protect\citeauthoryear{van der Wel et al.}{2021}]{van21} van der Wel A., Bezanson R., D'Eugenio F., Straatman C., Franx M., van Houdt J., Maseda M.~V., et al., 2021, ApJS, 256, 44

\bibitem[\protect\citeauthoryear{van der Wel et al.}{2016}]{van16} van der Wel A., Noeske K., Bezanson R., Pacifici C., Gallazzi A., Franx M., Mu{\~n}oz-Mateos J.~C., et al., 2016, ApJS, 223, 29

\bibitem[\protect\citeauthoryear{van der Wel et al.}{2014}]{van14} van der Wel A., Franx M., van Dokkum P.~G., Skelton R.~E., Momcheva I.~G., Whitaker K.~E., Brammer G.~B., et al., 2014, ApJ, 788, 28
  
\bibitem[\protect\citeauthoryear{van Dokkum et al.}{2013}]{van13} van Dokkum P.~G., Leja J., Nelson E.~J., Patel S., Skelton R.~E., Momcheva I., Brammer G., et al., 2013, ApJL, 771, L35
  
\bibitem[\protect\citeauthoryear{van Dokkum et al.}{2010}]{van10} van Dokkum P.~G., Whitaker K.~E., Brammer G., Franx M., Kriek M., Labb{\'e} I., Marchesini D., et al., 2010, ApJ, 709, 1018
  
\bibitem[\protect\citeauthoryear{Wang et al.}{2024}]{wan24} Wang J.-H., Li Z.-Y., Zhuang M.-Y., Ho L.~C., Lai L.-M., 2024, A\&A, 686, A100
  
\bibitem[\protect\citeauthoryear{Weaver et al.}{2022}]{wea22} Weaver J.~R., Kauffmann O.~B., Ilbert O., McCracken H.~J., Moneti A., Toft S., Brammer G., et al., 2022, ApJS, 258, 11
  
\bibitem[\protect\citeauthoryear{Whitaker et al.}{2017}]{whi17} Whitaker K.~E., Bezanson R., van Dokkum P.~G., Franx M., van der Wel A., Brammer G., F{\"o}rster-Schreiber N.~M., et al., 2017, ApJ, 838, 19

\bibitem[\protect\citeauthoryear{Whitney et al.}{2021}]{whi21} Whitney A., Ferreira L., Conselice C.~J., Duncan K., 2021, ApJ, 919, 139

\bibitem[\protect\citeauthoryear{Williams et al.}{1996}]{wil96} Williams R.~E., Blacker B., Dickinson M., Dixon W.~V.~D., Ferguson H.~C., Fruchter A.~S., Giavalisco M., et al., 1996, AJ, 112, 1335
  
\bibitem[\protect\citeauthoryear{Wilman et al.}{2020}]{wil20} Wilman D.~J., Fossati M., Mendel J.~T., Saglia R., Wisnioski E., Wuyts S., F{\"o}rster Schreiber N., et al., 2020, ApJ, 892, 1

\bibitem[\protect\citeauthoryear{Windhorst et al.}{2002}]{win02} Windhorst R.~A., Taylor V.~A., Jansen R.~A., Odewahn S.~C., Chiarenza C.~A.~T., Conselice C.~J., de Grijs R., et al., 2002, ApJS, 143, 113
  
\bibitem[\protect\citeauthoryear{Wright et al.}{2024}]{wri24} Wright L., Whitaker K.~E., Weaver J.~R., Cutler S.~E., Wang B., Carnall A., Suess K.~A., et al., 2024, ApJL, 964, L10
  
\bibitem[\protect\citeauthoryear{Wuyts et al.}{2013}]{wuy13} Wuyts S., F{\"o}rster Schreiber N.~M., Nelson E.~J., van Dokkum P.~G., Brammer G., Chang Y.-Y., Faber S.~M., et al., 2013, ApJ, 779, 135
  
\bibitem[\protect\citeauthoryear{Wuyts et al.}{2011}]{wuy11} Wuyts S., F{\"o}rster Schreiber N.~M., van der Wel A., Magnelli B., Guo Y., Genzel R., Lutz D., et al., 2011, ApJ, 742, 96

\bibitem[\protect\citeauthoryear{Xu \& Peng}{2024}]{xu24} Xu B., Peng Y., 2024, ApJ, 963, 15
  
\bibitem[\protect\citeauthoryear{Zamojski et al.}{2007}]{zam07} Zamojski M.~A., Schiminovich D., Rich R.~M., Mobasher B., Koekemoer A.~M., Capak P., Taniguchi Y., et al., 2007, ApJS, 172, 468

\bibitem[\protect\citeauthoryear{Zhuang, Li, \& Shen}{2024}]{zhu24} Zhuang M.-Y., Li J., Shen Y., 2024, ApJ, 962, 93

\bibitem[\protect\citeauthoryear{Zolotov et al.}{2015}]{zol15} Zolotov A., Dekel A., Mandelker N., Tweed D., Inoue S., DeGraf C., Ceverino D., et al., 2015, MNRAS, 450, 2327

  
  
  

  

\end{thebibliography}






\bsp	
\label{lastpage}
\end{document}